\documentclass[10pt,twocolumn,twoside]{opticajnl}
\journal{opticajournal} 

\makeatletter
\let\optica@maketitle\@maketitle
\renewcommand{\@maketitle}{%
  \vspace*{-0.45in}%
  \optica@maketitle
}
\makeatother

\setboolean{shortarticle}{false}

\usepackage{lineno}

\title{Implicit Neural Speckle Denoising}

\author[1,2,*]{Matthew R. Ziemann}
\author[3]{Casey J. Pellizzari}
\author[3]{Tyler J. Hardy}
\author[2]{Christopher A. Metzler}

\affil[1]{DEVCOM Army Research Laboratory, 2800 Powder Mill Rd, Adelphi, MD, USA}
\affil[2]{Department of Computer Science, University of Maryland, 8125 Paint Branch Dr, College Park, MD, USA}
\affil[3]{Department of Physics and Meteorology, United States Air Force Academy, USAFA, CO, USA}

\affil[*]{matthew.r.ziemann.civ@army.mil}

\affil[ ]{\rule{0pt}{1.5em}\textbf{Distribution Statement A.} Approved for public release: distribution is unlimited.}

\begin{abstract}
Speckle fundamentally limits coherent imaging by introducing multiplicative, spatially correlated noise that obscures scene structure. Removing speckle noise from dynamic scenes---that do not benefit from conventional speckle averaging---is particularly challenging. We introduce a training-free framework that combines a spatiotemporal implicit neural representation with an aperture-aware maximum-likelihood formulation to recover dynamic, speckle-free imagery directly from noisy observations. The coherent likelihood explicitly models the aperture-dependent spatial covariance of speckle, enabling adaptation to arbitrary pupil geometries without retraining. A matrix-free implementation based on FFT-accelerated operators, stochastic approximations, and conjugate gradients makes optimization practical for realistic image sizes. Meanwhile, a blind holdout criterion provides automatic early stopping without clean reference data. Simulated and laboratory results demonstrate improved spatial fidelity, temporal consistency, and robustness to varying speckle statistics relative to classical, unsupervised, and supervised baselines. 
\end{abstract}

\doi{}
\dates{}

\setboolean{displaycopyright}{false} 

\begin{document}

\maketitle

\begin{figure}[h]
    \centering
    \includegraphics[width=0.85\linewidth]{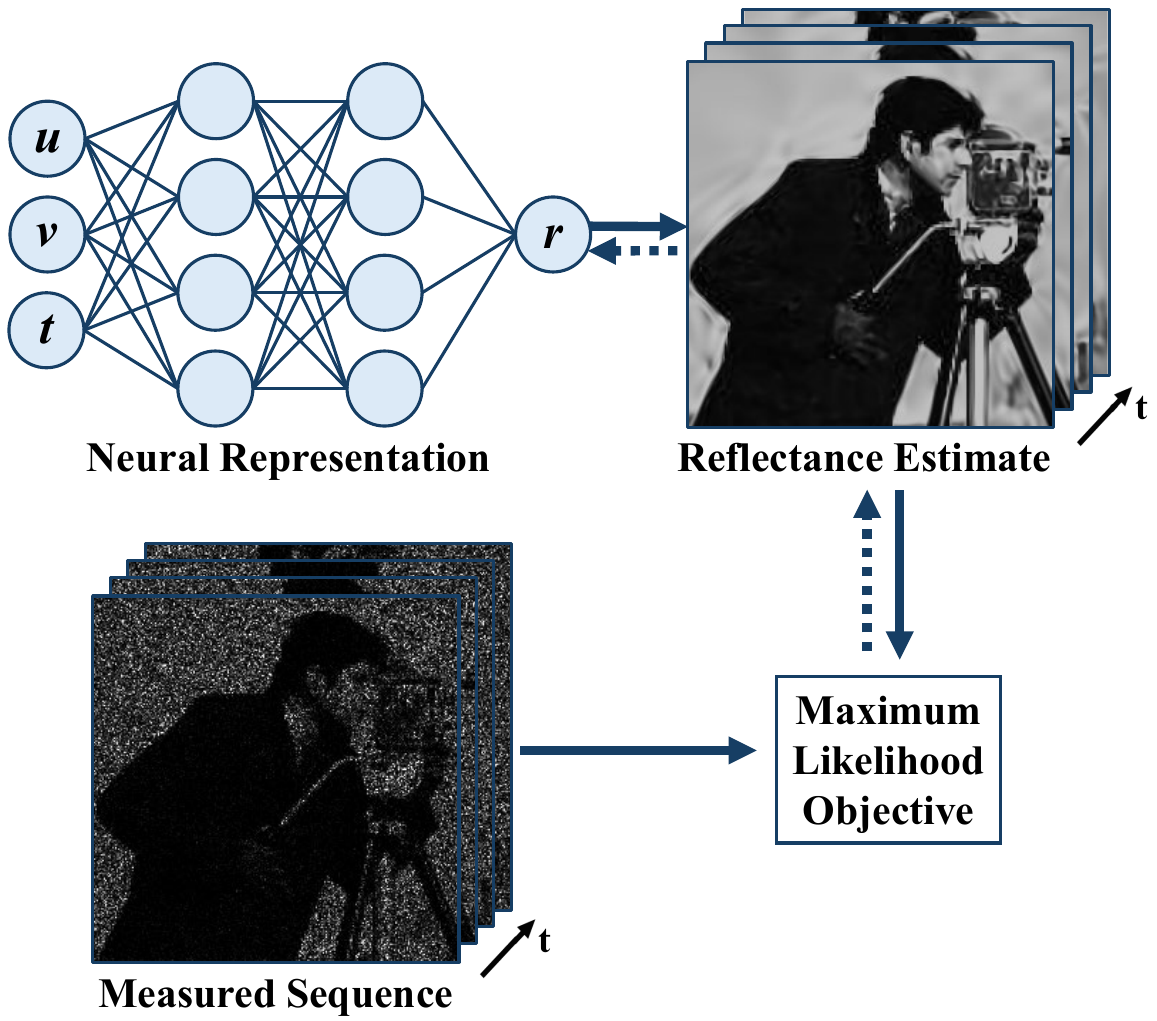}
    \caption{\textbf{Implicit neural speckle denoising.} A spatiotemporal INR is optimized directly from noisy coherent measurements using an aperture-aware maximum-likelihood objective, enabling dynamic speckle denoising without prior training.}
    \label{fig:overview}
\end{figure}

\section{Introduction}
\label{sec:intro}

Coherent imaging systems---including synthetic aperture radar (SAR), digital holography, and optical coherence tomography---measure the amplitude and phase of backscattered fields~\cite{Goodman2005,Goodman2007}. A fundamental consequence of coherent detection is speckle: a granular, multiplicative noise pattern caused by random interference from unresolved scatterers~\cite{Goodman2007}. Speckle obscures fine structure and hinders downstream tasks such as segmentation, classification, and change detection~\cite{Argenti2013,Kumar2020}, making speckle suppression a central challenge across coherent imaging modalities~\cite{Bianco2018}.

Existing approaches each resolve only part of the problem. Single-frame filters, from local-statistics methods such as Lee and Frost filtering~\cite{Lee1980,Frost1982} to wavelet and nonlocal approaches~\cite{Argenti2013,Parrilli2012}, reduce speckle by imposing spatial assumptions on a single realization, which generally trades noise suppression against spatial resolution. Multi-frame averaging and model-based estimation can improve this tradeoff by exploiting independent speckle realizations~\cite{Goodman2007,Pellizzari2024}, and recent aperture-aware models show the importance of retaining the coherent transfer function rather than assuming pixelwise independent speckle~\cite{Allen2025CLAMP}. However, these methods typically assume a static scene, and scene motion can produce artifacts.


Learning-based denoisers offer strong image priors, but coherent imaging rarely provides clean training targets: every measurement is speckled. Supervised networks sidestep this with synthetic data at the cost of domain mismatch~\cite{Ma2020,swin-unet}, and Noise2Noise-style training instead requires paired, registered scenes with independent speckle realizations~\cite{Lehtinen2018}. Self-supervised networks avoid clean targets but often require spatially decorrelated speckle and collections of representative measurements~\cite{Molini2022}. While some approaches bypass these requirements---sometimes needing only a single measurement~\cite{Dalsasso2022}---they supervise against per-pixel marginal statistics without explicitly modeling aperture-induced covariance, and remain specific to a given sensor and imaging mode. Training-free approaches such as Deep Image Prior and its extensions remove the data requirement by fitting a network directly to the corrupted measurement~\cite{DIP,Yoo2021,DVP}, but have not been adapted to the multiplicative, spatially correlated statistics of coherent speckle. None of these families simultaneously accommodates dynamic scenes and aperture-dependent speckle correlations without retraining.

Implicit neural representations (INRs) provide a natural neural prior for dynamic coherent imaging. By mapping continuous coordinates to signal values, INRs can compactly represent spatial and temporal structure while favoring coherent scene explanations over unstructured noise~\cite{deepsdf,nerf,siren,Tancik2020}. They have been successful across computational imaging tasks, including structured illumination microscopy~\cite{cao2024neural}, wavefront shaping~\cite{NeuWS}, intensity diffraction tomography~\cite{liu2022recovery}, and Fourier ptychographic microscopy~\cite{zhou2023fourier}, but, to our knowledge, they have not yet been coupled to an aperture-aware speckle likelihood for dynamic denoising.

In this work, we introduce a training-free framework for denoising dynamic coherent imagery, shown in Fig.~\ref{fig:overview}. The method uses a spatiotemporal INR as a prior over the underlying speckle-free scene, but fits that prior with the statistics of coherent imaging rather than with a generic image-domain loss. The resulting aperture-aware likelihood preserves the spatial covariance induced by the coherent transfer function, allowing known pupil geometries to be incorporated directly instead of approximating speckle as independent pixelwise noise. To make this dense Gaussian likelihood practical, we propose a matrix-free implementation that enables scaling to realistic image sizes. The same likelihood also provides a blind holdout stopping criterion, enabling per-sequence denoising without clean references, external training data, motion estimates, or aperture-specific retraining. Together, these components turn known coherent-imaging physics into a practical supervision signal for dynamic speckle suppression.

\begin{figure}[t]
    \centering
    \includegraphics[width=\linewidth]{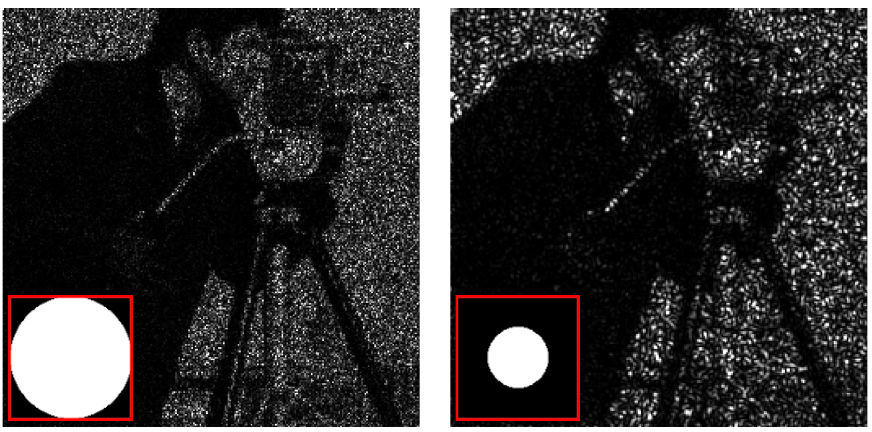}
    \caption{\textbf{Spatially correlated speckle.} Simulated coherent measurements generated using large (left) and small (right) aperture sizes at fixed sensor-plane sampling. Smaller relative apertures produce larger speckle grains and stronger spatial correlations, making denoising more challenging. Insets show the aperture in white and sampled frequency support in red.}
    \label{fig:speckle_example}
\end{figure}

\section{Likelihood-based implicit neural speckle denoising}
\label{sec:model}

\subsection{Aperture-aware coherent likelihood}
\label{subsec:speckle_stats}
Coherent images of rough objects are corrupted by speckle because many unresolved scatterers interfere within each resolution cell. Under the standard fully developed speckle model, the object-plane field at time $t$ can be written as $r_{t}^{1/2}\circ g_{t}$, where $r_{t}\in\mathbb{R}_+^N$ is the scene reflectance, $g_{t}\sim\mathcal{CN}(0,I)$ is a circular complex Gaussian speckle field, and $\circ$ denotes elementwise multiplication~\cite{Goodman2005}. We model coherent propagation through the sampled pupil as
\begin{equation}
    A = F^H \mathcal{D}(p) F,
    \label{eq:A_definition}
\end{equation}
where $F$ is the unitary two-dimensional DFT, $p$ is the real-valued pupil transmission function, and $\mathcal{D}(\cdot)$ forms a diagonal matrix. This model assumes negligible time-varying phase errors. Let $n_{t}\sim\mathcal{CN}(0,\sigma^2I)$ denote additive circular complex Gaussian measurement noise. The measured complex field is then
\begin{equation}
    \tilde{y}_{t} 
    = A \bigl(r_{t}^{1/2} \circ g_{t} \bigr) + n_{t},
    \label{eq:full_forward_model}
\end{equation}
which gives a probabilistic forward model for dynamic coherent measurements.

Conditioned on the reflectance $r_{t}$, the measurement is zero-mean complex Gaussian with covariance
\begin{equation}
    \Sigma_{t} 
    = A \mathcal{D}(r_{t}) A^H + \sigma^2 I.
    \label{eq:covariance}
\end{equation}
This covariance is the mechanism by which aperture-dependent speckle correlations enter the likelihood. If the pupil fills the sampled spatial-frequency support, the coherent operator is approximately unitary and the speckle is approximately conditionally independent across pixels. When the pupil occupies only part of the sampled frequency grid, the coherent point-spread function broadens and each speckle grain spans multiple samples, producing the spatial correlations illustrated in Fig.~\ref{fig:speckle_example}.

Assuming independent speckle and additive-noise realizations over time, the negative log-likelihood of a sequence of $T$ measurements, up to additive constants, is
\begin{equation}
    \mathcal{L}_{\text{MLE}}
    = \sum_{i=1}^T 
    \Big[
        \log \det \bigl(\Sigma_{t_i}\bigr)
        + \tilde{y}_{t_i}^H \Sigma_{t_i}^{-1} \tilde{y}_{t_i}
    \Big].
    \label{eq:full_nll}
\end{equation}
This likelihood provides the data-fidelity term used throughout the paper; unlike pixelwise speckle losses, it retains the aperture-dependent covariance induced by coherent propagation. Since minimizing this objective yields the maximum-likelihood estimate of the reflectance parameters, we refer to it as the MLE loss throughout the paper.  A full likelihood derivation is provided in Supplement~1.

\begin{figure}[t]
    \centering
    \includegraphics[width=\linewidth]{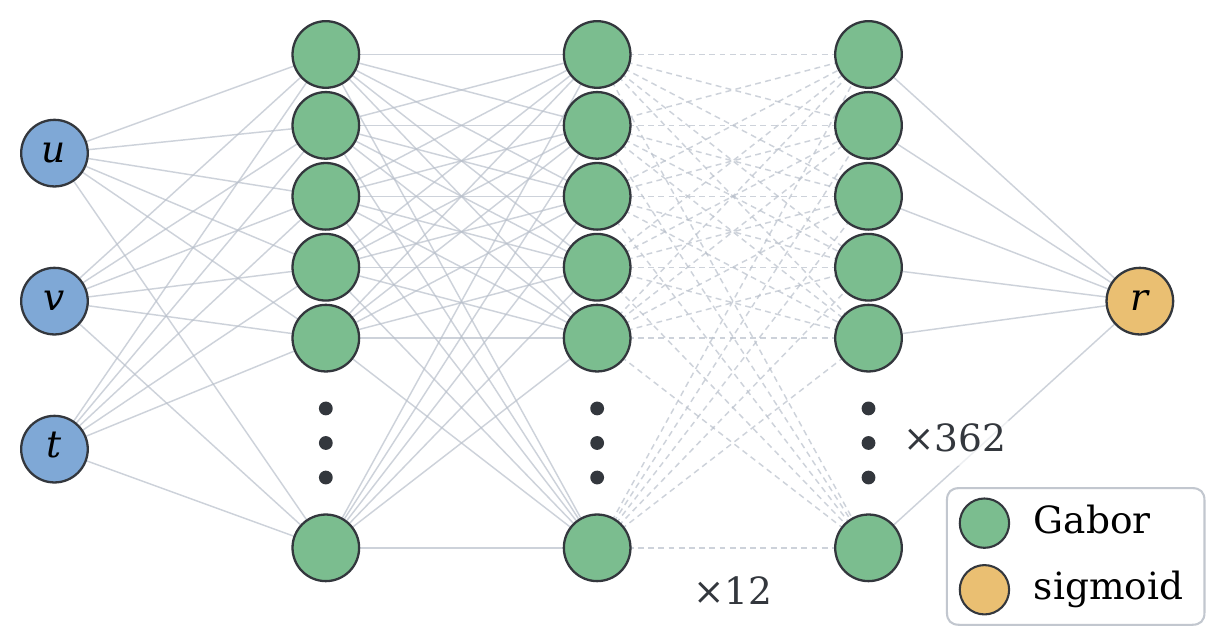}
    \caption{\textbf{Denoising neural architecture.} Diagram of the modified wavelet implicit neural representation (WIRE~\cite{wire}) used to optimize the likelihood-based loss for denoising.}
    \label{fig:WIRE}
\end{figure}

\subsection{Spatiotemporal implicit neural reconstruction}
\label{subsec:INR}
We estimate the dynamic reflectance sequence by representing it with an implicit neural representation (INR), a coordinate-based network that maps normalized spatial and temporal coordinates to reflectance:
\begin{equation}
    \hat r(u,v,t) = f_\theta(u,v,t),
    \qquad (u,v,t)\in[-1,1]^3.
\end{equation}
For each measurement sequence, we optimize the network parameters $\theta$ directly with the MLE loss in \eqref{eq:full_nll}. The resulting reconstruction is training-free: no clean reference images, external datasets, or pretrained denoising priors are required.

We implement $f_\theta$ using a wavelet implicit neural representation (WIRE)~\cite{wire}. WIRE is a fully connected multilayer perceptron that leverages complex Gabor nonlinearities, enabling localized oscillatory features that are well suited for representing fine spatial detail while maintaining stable optimization. For each hidden layer $\ell$, its input feature vector $h_{\ell-1}$ is passed through a linear transform followed by the complex Gabor activation
\begin{equation}
    h_\ell =
    \exp\!\left(
    j\,\omega_0 (W_\ell h_{\ell-1} + b_\ell)
    -
    \left|s_0 (W_\ell h_{\ell-1} + b_\ell)\right|^2
    \right),
\end{equation}
where $W_\ell$ and $b_\ell$ are layer weights and biases, $\omega_0$ controls the oscillation frequency, and $s_0$ controls the Gaussian envelope width. These parameters determine the frequency bias of the INR and are important for denoising, since fitting high-frequency structure too rapidly can cause the network to explain speckle before it has recovered the underlying reflectance.

In this work, we use a WIRE network with 12 hidden layers, each with 362 complex-valued neurons, totaling 1.6 million trainable parameters. This architecture is illustrated in Fig.~\ref{fig:WIRE}. We use an oscillation frequency $\omega_0=1$ and Gaussian envelope width $s_0=2$. These values are significantly lower than the published WIRE variant, necessary to encourage a stronger low-frequency fitting bias and a smoother transition to higher spatial frequencies during optimization. This allows the model to recover fine scene detail with fewer artifacts, with a comparison available in Section~3~\ref{subsec:ablation}. The output layer then linearly projects the complex-valued hidden representation to a single complex value, takes its real part, and applies a sigmoid nonlinearity to enforce nonnegativity of the resulting reflectance estimate. The network is fit to 32-frame windows and optimized using one randomly sampled frame per minibatch using Adam with learning rate $5\times10^{-5}$.

\begin{figure}[t]
    \centering
    \includegraphics[width=0.9\linewidth]{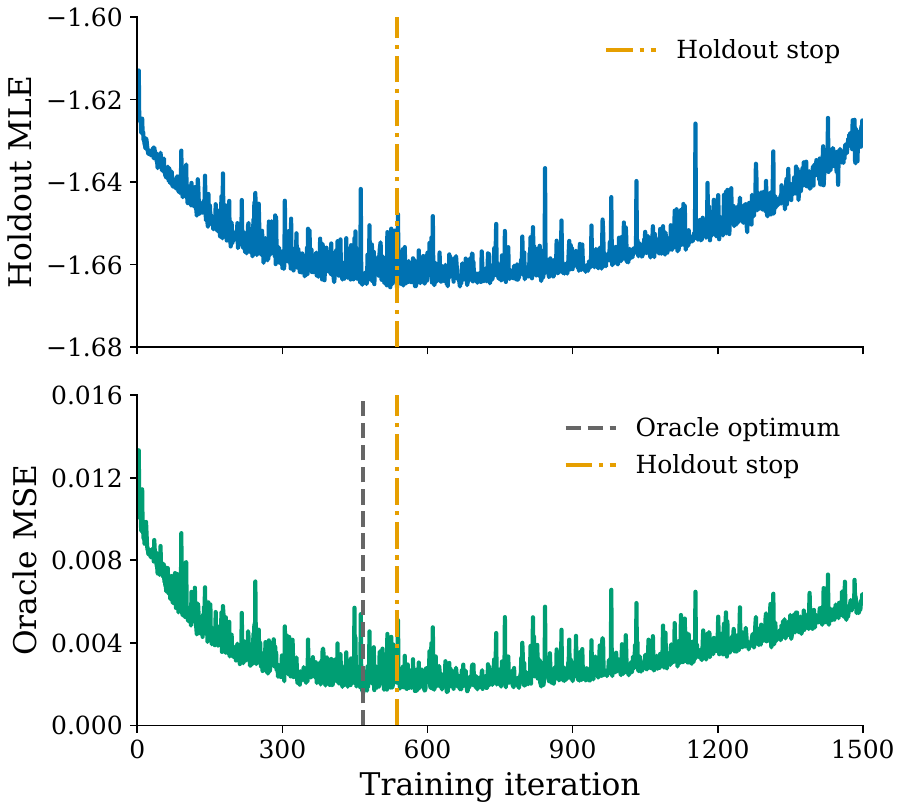}
    \caption{\textbf{Blind early stopping from the holdout likelihood.} The holdout MLE (top) tracks the unobservable ground-truth MSE (bottom) during INR optimization, reaching its minimum near the best reconstruction point while using two withheld noisy measurements.}
    \label{fig:holdout}
\end{figure}

\subsection{Blind holdout early stopping}
\label{subsec:holdout_early_stopping}
Like other neural denoisers optimized directly on noisy measurements, the proposed INR can eventually overfit noise within a sequence. This creates a practical model-selection problem: the likelihood objective is necessary for fitting the reconstruction, but unconstrained optimization can begin to explain frame-specific speckle rather than only the shared scene reflectance. We address this by using the same physics-based MLE loss as a blind early-stopping signal.

Specifically, we withhold two measurement frames from the optimization loss and periodically evaluate the MLE loss on those held-out measurements, stopping when the holdout MLE reaches its minimum. Because the INR represents a shared spatiotemporal reflectance field, improved scene reconstruction reduces the holdout loss. In contrast, overfitting independent frame-specific speckle in the optimized frames increases the holdout loss. Thus, the holdout loss provides a reference-free proxy for the otherwise unobservable reconstruction error.

In simulation, this behavior can be evaluated directly: the holdout MLE reaches its minimum near the oracle MSE minimum, providing a practical stopping signal without clean references or oracle access to ground truth (Fig.~\ref{fig:holdout}). This is an important advantage of the proposed likelihood-based formulation: the same coherent forward model used to fit the INR also provides a natural stopping criterion. 

All results reported for our proposed method use blind holdout early stopping unless otherwise stated. For the simulated data, we trigger early stopping on the minimum holdout MLE with a 150 iteration patience. The laboratory data exhibit noisier fitting, so we smooth the holdout MLE over 15 iterations and increase patience to 750 iterations.

\begin{figure*}[t]
    \centering
    \includegraphics[width=\linewidth]{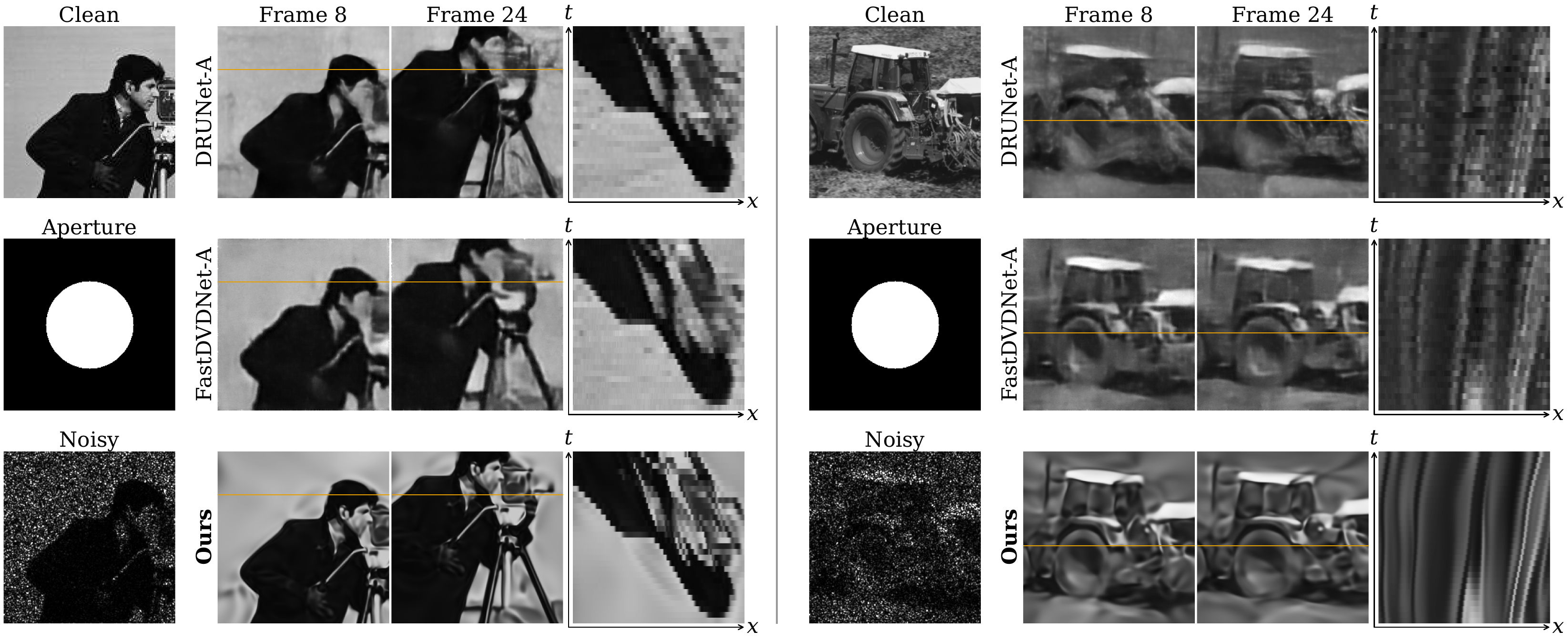}
    \caption{\textbf{Temporally consistent denoising.} Simulated small-aperture speckle denoising on translating \texttt{cameraman} (left) and DERF-HD \texttt{tractor} (right) sequences. The $x$--$t$ plots stack intensities along the indicated horizontal line across time. Our method recovers greater detail with more stable temporal trajectories. Supplemental videos more clearly illustrate denoising performance.}
    \label{fig:sim_05A_qualitative}
\end{figure*}

\begin{figure*}[t]
    \centering
    \includegraphics[width=\linewidth]{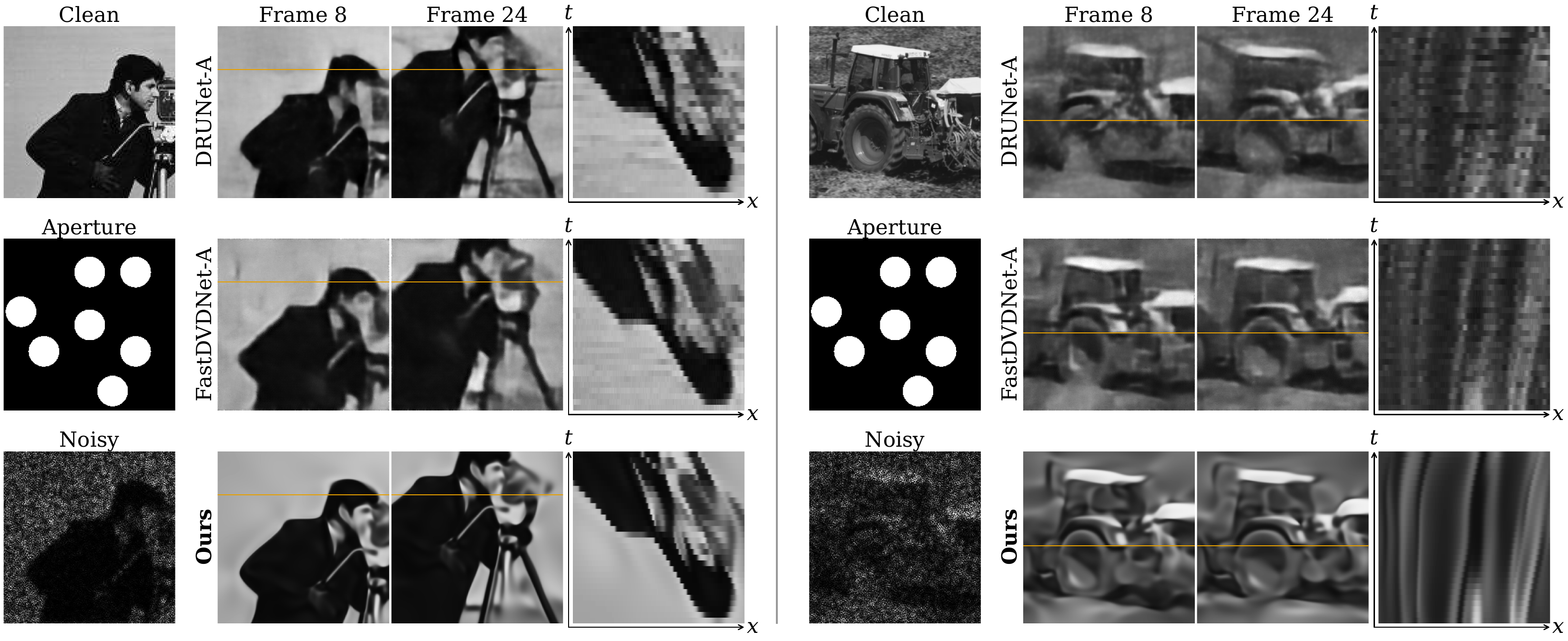}
    \caption{\textbf{Aperture-aware denoising.} Simulated Golay-aperture speckle denoising on translating \texttt{cameraman} (left) and DERF-HD \texttt{tractor} (right) sequences. Our method recovers greater detail with more stable temporal trajectories. Supplemental videos more clearly illustrate denoising performance.}
    \label{fig:sim_golay_qualitative}
\end{figure*}

\begin{table}[b]
\centering
\caption{\textbf{Runtime scaling of likelihood-based denoising.} Average denoising runtime per frame for explicit covariance evaluation and the proposed matrix-free implementation.}
\label{tab:mle_runtime_scaling}
\footnotesize
\setlength{\tabcolsep}{4pt}
\begin{tabular}{l|cc}
\hline
Resolution & Explicit MLE (s) & Matrix-free MLE (s) \\
\hline
$64\times64$   & 32 & 41 \\
$128\times128$ & 963 & 47 \\
$256\times256$ & OOM$^\dagger$  & 85 \\
$512\times512$ & OOM$^\dagger$  & 222 \\
\hline
\multicolumn{3}{l}{\footnotesize $^\dagger$Out of memory on 96 GB VRAM GPU}
\end{tabular}
\end{table}

\subsection{Matrix-free likelihood optimization}
\label{sec:loss}
The aperture-aware likelihood in \eqref{eq:full_nll} is the central data-fidelity term in the proposed method, but direct evaluation is not computationally practical at realistic image sizes. For each frame, explicit evaluation requires forming the dense covariance matrix $\Sigma_{t_i}\in\mathbb{C}^{N\times N}$, computing its log determinant, and evaluating an inverse quadratic form. This gives $\mathcal{O}(N^3)$ time and $\mathcal{O}(N^2)$ memory per frame, which quickly dominates INR optimization.

We instead evaluate the same likelihood matrix-free, without explicitly forming $\Sigma_{t_i}$ or its inverse. The implementation exploits three ingredients: \emph{(i)} FFT-based application of the coherent covariance, \emph{(ii)} stochastic Lanczos quadrature (SLQ) for the log-determinant, and \emph{(iii)} conjugate gradients (CG) for the inverse quadratic form. Each covariance application is expressed through the coherent forward model and implemented with two-dimensional FFTs, while SLQ and CG require only repeated covariance matrix--vector products. The resulting per-frame cost scales as $\mathcal{O}((SK+G)N\log N)$, where $S$ is the number of Hutchinson probes, $K$ is the number of Lanczos iterations, and $G$ is the number of CG iterations. In practice, all presented experiments use $S=8$ Hutchinson probes, $K=20$ Lanczos iterations, and at most $G=200$ CG iterations.

Table~\ref{tab:mle_runtime_scaling} compares denoising times for both explicit and matrix-free MLE loss implementations for 400 optimization iterations. The explicit covariance implementation remains feasible at $64\times64$ resolution, but becomes approximately 20 times slower by $128\times128$ and runs out of memory at $256\times256$ on a 96 GB GPU. In contrast, the matrix-free implementation remains feasible through $512\times512$, enabling likelihood-based INR optimization at the resolutions used in our experiments. Matrix-free derivations and accuracy comparisons are provided in Supplement~1.

\section{Results}
\label{sec:results}

\subsection{Baselines and evaluation metrics}
\label{subsec:baseline_comparisons}

Across simulated and laboratory experiments, we compare against classical, unsupervised, and supervised denoising baselines: H-BM3D~\cite{BM3D,mulog}, Deep Video Prior (DVP)~\cite{DVP}, DRUNet~\cite{drunet}, and FastDVDNet~\cite{fastdvdnet}. The supervised models are evaluated both with a standard pixelwise-speckle training model (corresponding to uncorrelated speckle) and with aperture-aware retraining, denoted by ``-A,'' in which training data are generated using the same aperture model as the test measurements. For laboratory experiments, this aperture model is estimated from the measured pupil support and is the same one used in the proposed MLE loss. Together, these baselines provide representative comparisons across classical single-frame despeckling, unsupervised neural video reconstruction, supervised image and video denoising, and aperture-matched retraining for spatially correlated speckle.

We report PSNR, SSIM~\cite{ssim}, and ST-RRED~\cite{strred} relative to clean simulated references when available. DVP is given oracle early stopping to report its best attainable reconstruction, while our method uses the blind holdout rule described in Section~2~\ref{subsec:holdout_early_stopping}. Baseline implementations, training protocols, and metric details are provided in Supplement~1.

\begin{table}[tbh]
\centering
\caption{\textbf{Denoising performance on simulated speckle.} Average PSNR (dB), SSIM, and ST-RRED for DERF-HD videos under two circular aperture sizes at fixed sensor-plane sampling. Best results are in bold.}
\label{tab:denoising_aperture_results}
\scriptsize
\setlength{\tabcolsep}{2pt}
\resizebox{\linewidth}{!}{%
\begin{tabular}{l|ccc|ccc}
\hline
& \multicolumn{3}{c|}{Large Aperture}
& \multicolumn{3}{c}{Small Aperture} \\
Method
& PSNR $\uparrow$ & SSIM $\uparrow$ & ST-RRED $\downarrow$
& PSNR $\uparrow$ & SSIM $\uparrow$ & ST-RRED $\downarrow$ \\
\hline
Noisy input              & 10.6 & 0.10 & 1933 & 10.7 & 0.10 & 2671 \\
H-BM3D                   & 22.7 & 0.57 & 1264 & 16.3 & 0.28 & 2867 \\
DVP                      & 16.1 & 0.38 & 2811 & 16.2 & 0.38 & 3366 \\
DRUNet                   & 24.3 & 0.66 & 766 & 17.1 & 0.28 & 2588 \\
DRUNet-A$^\dagger$       & 24.3 & 0.66 & 856 & 21.9 & 0.55 & 1828 \\
FastDVDNet               & 24.3 & 0.63 & 628 & 19.3 & 0.40 & 2599 \\
FastDVDNet-A$^\dagger$   & 24.7 & 0.64 & 505 & 22.8 & 0.55 & 1084 \\
\textbf{Ours}            & \textbf{25.6} & \textbf{0.72} & \textbf{201}
                         & \textbf{22.9} & \textbf{0.59} & \textbf{574} \\
\hline
\end{tabular}}
\vspace{1pt}

\raggedright\scriptsize $^\dagger$Aperture-aware (-A) supervised models were retrained separately for each aperture size.
\end{table}

\subsection{Simulated experiments}

We first evaluate dynamic denoising on simulated coherent measurements from the DERF-HD benchmark~\cite{DERF-HD,derf} and a translating \texttt{cameraman} sequence. We use the first 32 frames of each sequence with source frames resized to $256\times256$ resolution. Speckled complex-field measurements are generated using \eqref{eq:full_forward_model} with additive noise level $\sigma=0.02$ under two circular apertures at fixed sensor-plane sampling: a large aperture that fills the sampled Fourier grid and a smaller aperture with half the diameter that produces larger speckle grains and stronger spatial speckle correlation, as illustrated in Fig.~\ref{fig:speckle_example}. Additional dataset details are provided in Supplement~1.

Table~\ref{tab:denoising_aperture_results} reports average performance across DERF-HD. For large-aperture speckle, our method achieves the best average performance across all metrics, reaching 25.6 dB PSNR, 0.72 SSIM, and 201 ST-RRED. This improves both PSNR and SSIM compared to the strongest supervised baselines, and reduces ST-RRED by more than a factor of two. The remaining baselines improve over the noisy input but leave substantial artifacts and significant temporal irregularity.

The robustness advantage is more pronounced in the small-aperture setting, where stronger speckle correlations cause standard supervised denoisers to degrade sharply: DRUNet drops from 24.3 to 17.1 dB, and FastDVDNet drops from 24.3 to 19.3 dB. Aperture-aware retraining recovers much of this loss, confirming the importance of matching the speckle statistics at training time, but it requires specialized training data for each aperture condition. In contrast, our method incorporates the known aperture directly in the likelihood model and achieves improved PSNR, SSIM, and ST-RRED without aperture-specific retraining.

Figure~\ref{fig:sim_05A_qualitative} provides examples of simulated small-aperture denoising on the translating \texttt{cameraman} and DERF-HD \texttt{tractor} sequences. Our reconstruction preserves sharper, temporally stable structure across both sequences. Notably, the supervised baselines exhibit significant frame-to-frame feature variations. These differences are more pronounced in video format in Visualizations~1--2.

To explore the performance of our method on more challenging apertures, we simulate speckle with a sparse, distributed aperture based on the Golay 6+1 metalens~\cite{golay}. This is composed of a central sub-aperture surrounded by 6 off-axis sub-apertures. An example of this aperture and the resulting denoising performance can be seen in Fig.~\ref{fig:sim_golay_qualitative}. As with the large- and small-aperture case, our method continues to outperform all baselines by recovering greater detail with significantly reduced frame-to-frame variation. On DERF-HD, our method matches PSNR performance of the best-performing retrained baseline FastDVDNet-A at 22.2 dB, improves SSIM from 0.52 to 0.55, and improves ST-RRED from 1274 to 701. Denoising performance is more pronounced in video format in Visualizations~3--4.

\begin{table}[tbh]
\centering
\caption{\textbf{Architecture and loss ablations.} Comparison of network architectures and loss functions for the DERF-HD \texttt{sunflower} video with simulated large-aperture speckle. Our tuned WIRE and MLE loss outperform alternatives.}
\label{tab:architecture_loss_ablation}
\footnotesize
\setlength{\tabcolsep}{5pt}
\begin{tabular}{ll|ccc}
\hline
INR & Loss & PSNR $\uparrow$ & SSIM $\uparrow$ & ST-RRED $\downarrow$ \\
\hline
Tuned WIRE & MLE & \textbf{27.9} & \textbf{0.84} & \textbf{147} \\
Original WIRE & MLE & 23.7 & 0.65 & 704 \\
SIREN & MLE & 22.6 & 0.61 & 259 \\
MLP & MLE & 18.7 & 0.38 & 3556 \\
Tuned WIRE & MSE & 25.7 & 0.77 & 211 \\
Original WIRE & MSE & 22.3 & 0.61 & 678 \\
\hline
\end{tabular}
\end{table}

\begin{figure*}[t]
    \centering
    \includegraphics[width=0.7\linewidth]{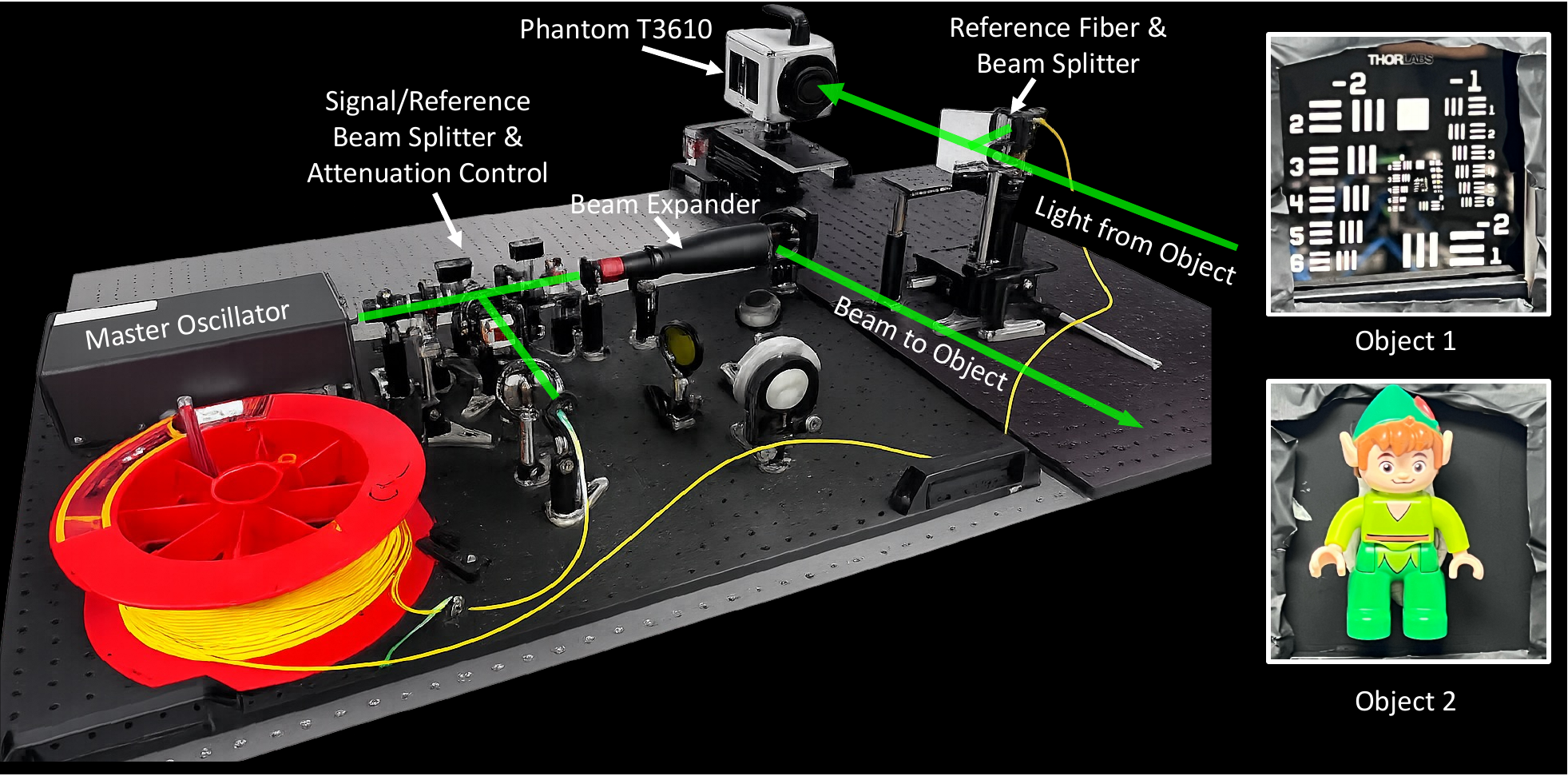}
    \caption{\textbf{Laboratory data collection.} Off-axis digital holography setup used to collect speckled videos of moving objects. A signal path flood-illuminates targets 2.6 m away, and the return interferes with a fiber-delivered reference beam at the camera.}
    \label{fig:exp_setup}
\end{figure*}

\begin{figure*}[tbh]
    \centering
    \includegraphics[width=0.8\linewidth]{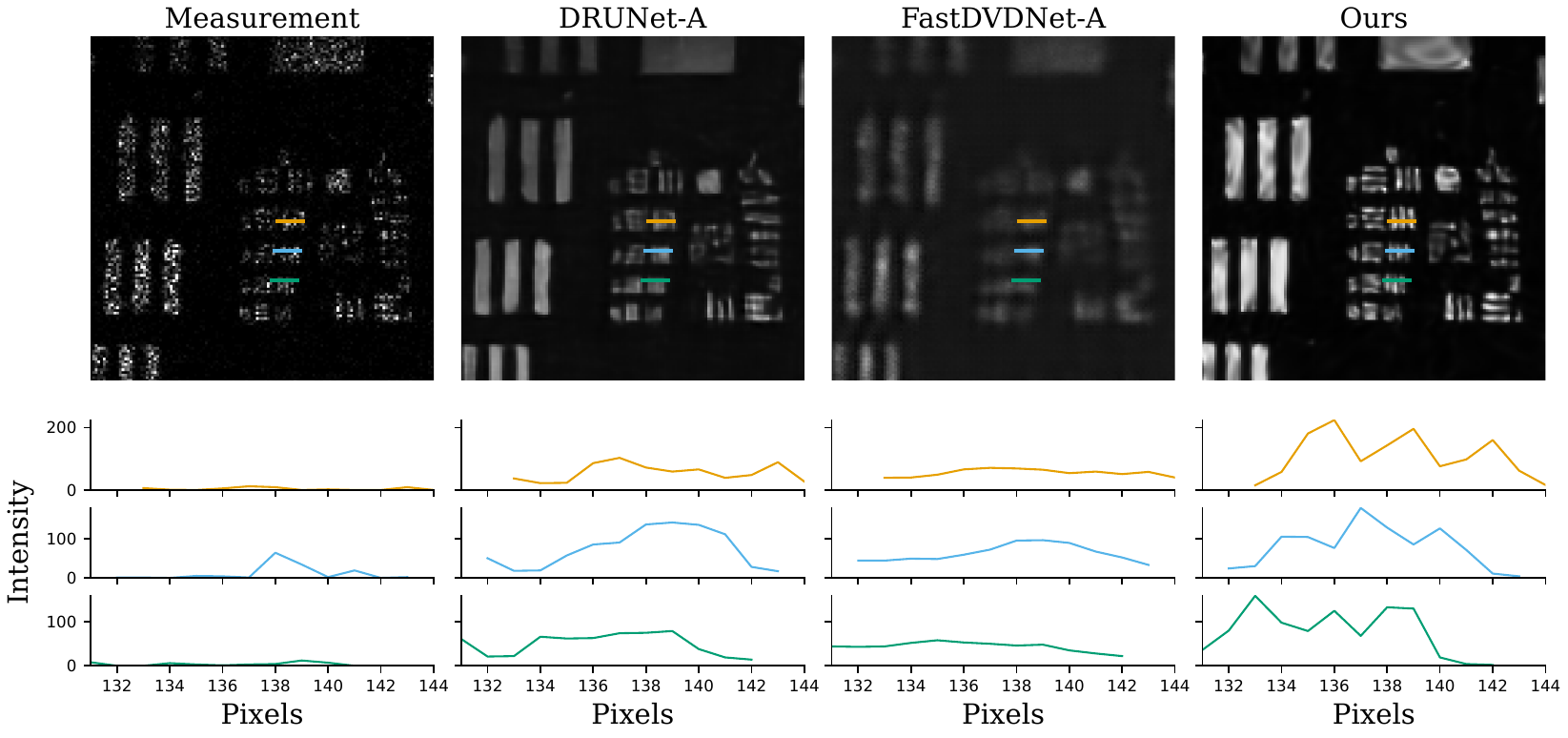}
    \caption{\textbf{Improved detail recovery on laboratory data.} Intensity profiles across laboratory USAF resolution bars. Our method preserves more distinct peaks and valleys, indicating improved recovery of fine spatial structure.}
    \label{fig:lab_peaks}
\end{figure*}

\subsection{Ablation experiments}
\label{subsec:ablation}
To separate the contributions of the INR architecture and the likelihood model, we ablate each on the $256\times256$ DERF-HD \texttt{sunflower} sequence with simulated large-aperture speckle, with results summarized in Table~\ref{tab:architecture_loss_ablation}. All variants in this comparison use oracle early stopping so that differences reflect the architecture and objective rather than the stopping rule.

We first fix the MLE loss and vary the INR backbone. A standard MLP recovers little fine detail, and SIREN~\cite{siren} introduces significant artifacts, while the tuned WIRE parameterization described in Section~2~\ref{subsec:INR} performs best. Relative to the original WIRE~\cite{wire} under the same MLE objective, tuning improves PSNR from 23.7 to 27.9~dB, SSIM from 0.65 to 0.84, and ST-RRED from 704 to 147. We then fix the tuned WIRE backbone and replace the aperture-aware MLE objective with an intensity-domain mean squared error (MSE) loss. This variant still benefits from the spatiotemporal INR prior, but removing the likelihood-based objective reduces denoising performance across all metrics. Together, these comparisons show that the observed gains come from both the tuned spatiotemporal INR prior and the likelihood model's treatment of aperture-dependent speckle correlations.

\begin{figure*}[t]
    \centering
    \includegraphics[width=\linewidth]{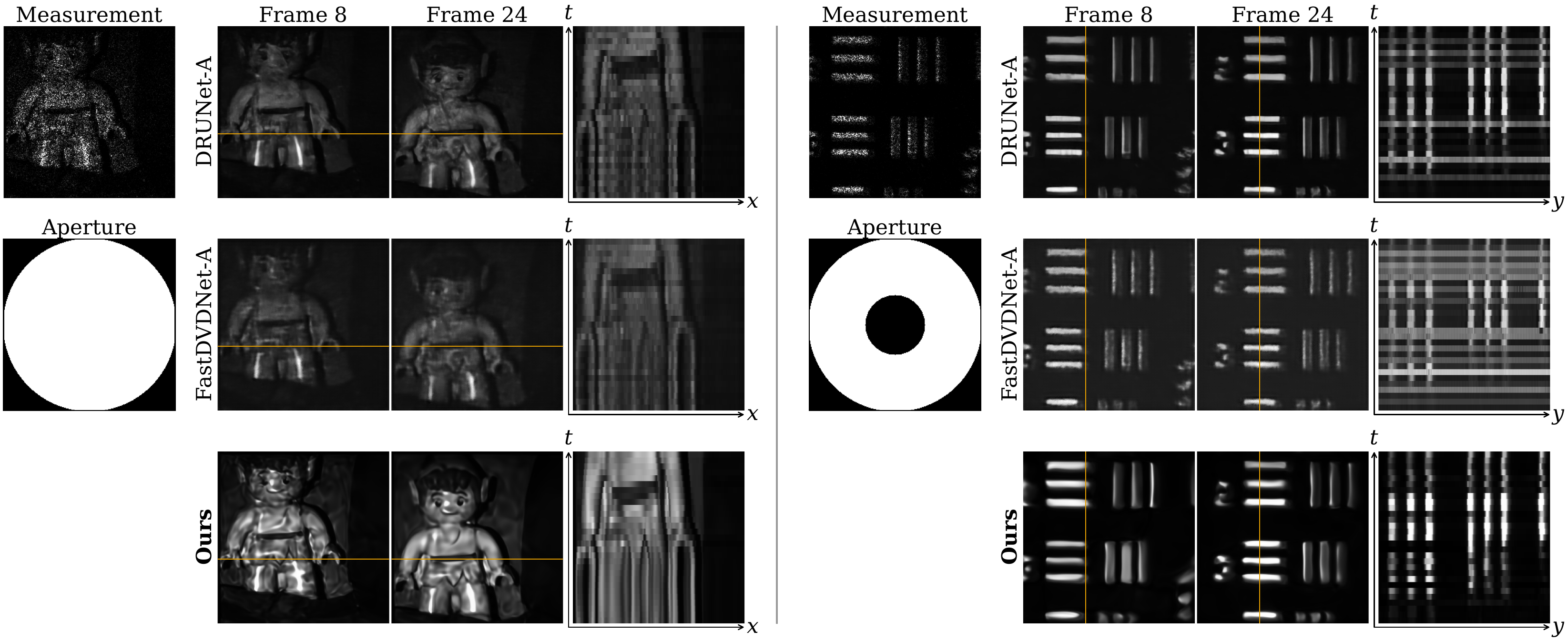}
    \caption{\textbf{Temporally consistent laboratory denoising.} Digital holography denoising of a translating toy target measured with a conventional circular aperture (left) and a translating resolution target measured with an annular aperture (right). Our method retains more temporally consistent features and adapts to the distinct aperture shapes without training data, while the retrained supervised baselines produce details that vary across frames. Supplemental videos more clearly illustrate denoising performance.}
    \label{fig:lab_qual}
\end{figure*}

\subsection{Laboratory experiments}

To validate our approach with real data, we collected off-axis digital holography videos of manually translated targets using the setup in Fig.~\ref{fig:exp_setup}, described in detail in Supplement~1. The measurements are demodulated to $256\times256$ complex-valued fields before denoising. Because clean reference images are unavailable for the laboratory measurements, we do not report full-reference metrics such as PSNR or SSIM. Instead, we evaluate laboratory performance using reference-free visual comparisons, temporal $x$--$t$ slices, and intensity-profile measurements. We limit these comparisons to the best-performing supervised baselines for space, with additional baselines shown in Visualizations~5--7.

We first evaluate a translating USAF resolution target, shown in Fig.~\ref{fig:lab_peaks}. All methods substantially reduce visible speckle, and DRUNet-A in particular produces competitive frame-wise reconstructions (though they vary from frame to frame). However, differences become apparent in the high-frequency resolution bars. The intensity profiles in Fig.~\ref{fig:lab_peaks} show that the supervised baselines distort the high-frequency bar structure, causing adjacent bars to merge together. Our method preserves distinct peaks and valleys corresponding to individual bars.

To test performance on a less structured laboratory target, we also collected a translating toy sequence. Figure~\ref{fig:lab_qual} (left) shows that DRUNet-A and FastDVDNet-A suppress speckle and produce plausible-looking individual frames, but some recovered features in the face and upper body shift, blur, or re-form across time. In contrast, the proposed likelihood-guided INR recovers sharper object structure while keeping features more stable throughout the sequence. The $x$--$t$ slices further highlight this behavior, showing more coherent trajectories corresponding to the translating object without sudden frame-to-frame variation. This indicates that the temporal consistency benefit is not restricted to periodic resolution bars.

We then evaluate aperture awareness in the laboratory by replacing the circular pupil with an annular aperture, with results shown in Fig.~\ref{fig:lab_qual} (right). This annular aperture changes the coherent transfer function, and therefore the spatial covariance of the measured speckle, rather than merely changing image content. This laboratory modification also changes the field of view and magnification, so the resulting video shows a more highly magnified region of the resolution target containing only coarser bar groups. For this experiment, we estimate the inner and outer radius of the pupil support from the Fourier-domain magnitude of the demodulated laboratory measurements and use the resulting annular mask as the aperture function $p$ in the likelihood model. The measured annular support is shown in Supplement~1. We find that our method produces temporally consistent denoising; although the absence of finer bar groups makes spatial-resolution differences less apparent than in Fig.~\ref{fig:lab_peaks}, our method preserves the coarse bar structure more consistently and reduces the frame-to-frame variation present in the supervised baselines. Together, the laboratory results show that the proposed likelihood-based INR extends beyond idealized simulations to dynamic digital holography measurements with real optical imperfections, varied scene structure, and changing aperture geometry.

\newpage
\section{Discussion and conclusion}

We introduced a likelihood-guided implicit neural denoising framework for dynamic coherent imaging. The method represents time-varying scene reflectance with a spatiotemporal INR and fits it directly to noisy complex-field measurements using an aperture-aware coherent likelihood. This formulation accounts for spatially correlated speckle induced by the coherent transfer function, adapts naturally to known aperture changes, and supports holdout-based stopping without clean references or simulated ground truth. A matrix-free implementation makes the likelihood feasible to evaluate at realistic image resolutions.

Across simulated dynamic scenes, the method improved spatial fidelity and temporal consistency relative to classical, unsupervised, and supervised baselines. Laboratory digital holography experiments further showed that the framework extends beyond idealized simulations, resolving finer spatial detail and producing temporally stable denoising under both circular and annular apertures. These results suggest that physics-based likelihoods can reduce dependence on external training data and aperture-specific retraining in dynamic coherent imaging.

The primary limitation of the current approach is computational cost: each sequence requires per-instance INR optimization and repeated likelihood evaluation. Although the matrix-free likelihood makes this optimization feasible at the image sizes considered here, it remains much slower than feed-forward supervised denoisers at test time. Future work could accelerate reconstruction by reusing INR initializations across neighboring sequences, parallelizing spatial patches, adopting coarse-to-fine solver schedules, or developing lighter-weight spatiotemporal INR architectures.

The statistical model also assumes fully developed speckle, where the desired reconstruction is an intensity-like reflectance and the instantaneous coherent field is treated as a nuisance random process. This is appropriate for the rough-object and diffuse coherent-imaging regimes studied here, but richer object models may be needed for scenes with deterministic phase structure, persistent specular components, partially developed speckle, or temporal speckle correlation. Extensions of the framework to mixed coherent--diffuse fields and correlated speckle likelihoods~\cite{Chen2026} are promising directions for future work.

Overall, these results demonstrate that combining aperture-aware coherent likelihood models with implicit neural representations provides a promising path toward physics-guided implicit reconstruction of dynamic coherent imagery, particularly when clean references, matched training data, or motion estimates are unavailable.

\begin{backmatter}
\bmsection{Funding} 
This work was supported in part by the Office of Naval Research (ONR) (N00014-23-1-2752); the National Science Foundation (NSF) (CAREER 2339616); and the Joint Directed Energy Transition Office (JDETO).

\bmsection{Disclosures} The authors declare no conflicts of interest.

\bmsection{Acknowledgment}
Preliminary portions of this work were presented at the SPIE Unconventional Imaging, Sensing, and Adaptive Optics in 2024, "Time-varying implicit neural representations for unsupervised speckle denoising in dynamic scenes" \cite{Ziemann2024}. This work was supported in part by high-performance computer time and resources from the DoD High Performance Computing Modernization Program. The authors thank Profs. Shirin Jalali and Arian Maleki for their input on implementation methodology and evaluation framework.

The U.S. Government is authorized to reproduce and distribute reprints for governmental purposes notwithstanding any copyright notation thereon. The views and conclusions contained herein are those of the authors and should not be interpreted as necessarily representing the official policies or endorsements, either expressed or implied, of the Dept. of Army or the Office of the Under Secretary of Defense for Research and Engineering (OUSD(R\&E)) or the U.S. Government. 

\bmsection{Data availability}
Data underlying the results presented in this paper are not publicly available at this time but may be obtained from the authors upon reasonable request.

\bmsection{Supplemental document}
See Supplement~1 for supporting content.
\end{backmatter}

\bibliography{refs}

@article{liu2022recovery,
  title={Recovery of continuous 3d refractive index maps from discrete intensity-only measurements using neural fields},
  author={Liu, Renhao and Sun, Yu and Zhu, Jiabei and Tian, Lei and Kamilov, Ulugbek S},
  journal={Nature Machine Intelligence},
  volume={4},
  number={9},
  pages={781--791},
  year={2022},
  publisher={Nature Publishing Group UK London}
}

@article{zhou2023fourier,
  title={Fourier ptychographic microscopy image stack reconstruction using implicit neural representations},
  author={Zhou, Haowen and Feng, Brandon Y and Guo, Haiyun and Lin, Siyu and Liang, Mingshu and Metzler, Christopher A and Yang, Changhuei},
  journal={Optica},
  volume={10},
  number={12},
  pages={1679--1687},
  year={2023},
  publisher={Optica Publishing Group}
}

@BOOK{Goodman2005,
  title     = "Introduction to Fourier Optics",
  author    = "Goodman, Joseph W",
  publisher = "W.H. Freeman",
  address   = "New York, NY",
  edition   =  3,
  month     =  jan,
  year      =  2005
}

@BOOK{Goodman2007,
  title     = "Speckle Phenomena in Optics: Theory and Applications",
  author    = "Goodman, Joseph W",
  publisher = "Roberts and Company Publishers",
  year      =  2007,
  language  = "en"
}

@ARTICLE{Yoo2021,
  title    = "Time-Dependent Deep Image Prior for Dynamic {MRI}",
  author   = "Yoo, Jaejun and Jin, Kyong Hwan and Gupta, Harshit and Yerly,
              Jerome and Stuber, Matthias and Unser, Michael",
  journal  = "IEEE Trans. Med. Imaging",
  volume   =  40,
  number   =  12,
  pages    = "3337--3348",
  month    =  dec,
  year     =  2021,
  language = "en"
}

@INPROCEEDINGS{DIP,
  title     = "Deep Image Prior",
  author    = "Lempitsky, Victor and Vedaldi, Andrea and Ulyanov, Dmitry",
  booktitle = "2018 IEEE/CVF Conference on Computer Vision and Pattern
               Recognition",
  publisher = "IEEE",
  pages     = "9446--9454",
  month     =  jun,
  year      =  2018
}

@ARTICLE{BM3D,
  title     = "Image denoising by sparse 3-{D} transform-domain collaborative
               filtering",
  author    = "Dabov, Kostadin and Foi, Alessandro and Katkovnik, Vladimir and
               Egiazarian, Karen",
  journal   = "IEEE Trans. Image Process.",
  publisher = "Institute of Electrical and Electronics Engineers (IEEE)",
  volume    =  16,
  number    =  8,
  pages     = "2080--2095",
  month     =  aug,
  year      =  2007,
  language  = "en"
}

@InProceedings{div2k,
	author = {Agustsson, Eirikur and Timofte, Radu},
	title = {{NTIRE} 2017 Challenge on Single Image Super-Resolution: Dataset and Study},
	booktitle = {The IEEE Conference on Computer Vision and Pattern Recognition (CVPR) Workshops},
	month = {July},
	year = {2017}
}

@ARTICLE{Pellizzari2024,
  title     = "Speckle-free coherent imaging through deep turbulence",
  author    = "Pellizzari, Casey J and Bate, Timothy J and Mandyam, Maya G and
               Radosevich, Cameron J and Horst, Samuel and Spencer, Mark F",
  journal   = "Optics Letters",
  publisher = "Optica Publishing Group",
  volume    =  49,
  number    =  12,
  pages     = "3508--3511",
  month     =  jun,
  year      =  2024
}

@ARTICLE{Pellizzari2017,
  title    = "Phase-error estimation and image reconstruction from
              digital-holography data using a Bayesian framework",
  author   = "Pellizzari, Casey J and Spencer, Mark F and Bouman, Charles A",
  journal  = "J. Opt. Soc. Am. A Opt. Image Sci. Vis.",
  volume   =  34,
  number   =  9,
  pages    = "1659--1669",
  month    =  sep,
  year     =  2017,
  language = "en"
}

@ARTICLE{Kumar2020,
  title     = "Speckle denoising techniques in imaging systems",
  author    = "Kumar, Manoj and Tounsi, Yassine and Kaur, Karmjit and Nassim,
               Abdelkrim and Mandoza-Santoyo, Fernando and Matoba, Osamu",
  journal   = "J. Opt.",
  publisher = "IOP Publishing",
  volume    =  22,
  number    =  6,
  pages     =  063001,
  month     =  jun,
  year      =  2020
}

@INPROCEEDINGS{deepsdf,
  title     = "{DeepSDF}: Learning continuous signed distance functions for
               shape representation",
  author    = "Park, Jeong Joon and Florence, Peter and Straub, Julian and
               Newcombe, Richard and Lovegrove, Steven",
  booktitle = "2019 IEEE/CVF Conference on Computer Vision and Pattern
               Recognition (CVPR)",
  publisher = "IEEE",
  pages     = "165--174",
  month     =  jun,
  year      =  2019
}

@ARTICLE{nerf,
  title     = "{NeRF}: representing scenes as neural radiance fields for view
               synthesis",
  author    = "Mildenhall, Ben and Srinivasan, Pratul P and Tancik, Matthew and
               Barron, Jonathan T and Ramamoorthi, Ravi and Ng, Ren",
  journal   = "Commun. ACM",
  publisher = "Association for Computing Machinery (ACM)",
  volume    =  65,
  number    =  1,
  pages     = "99--106",
  month     =  jan,
  year      =  2022,
  language  = "en"
}

@ARTICLE{siren,
  title   = "Implicit neural representations with periodic activation functions",
  author  = "Sitzmann, Vincent and Martel, Julien and Bergman, Alexander and
             Lindell, David and Wetzstein, Gordon",
  journal = "Adv. Neural Inf. Process. Syst.",
  volume  =  33,
  pages   = "7462--7473",
  year    =  2020
}

@ARTICLE{pytorch,
  title     = "{PyTorch}: An imperative style, high-performance deep learning
               library",
  author    = "Paszke, Adam and Gross, Sam and Massa, Francisco and Lerer, Adam
               and Bradbury, James and Chanan, Gregory and Killeen, Trevor and
               Lin, Zeming and Gimelshein, Natalia and Antiga, Luca and
               Desmaison, Alban and Kopf, Andreas and Yang, Edward and DeVito,
               Zachary and Raison, Martin and Tejani, Alykhan and Chilamkurthy,
               Sasank and Steiner, Benoit and Fang, Lu and Bai, Junjie and
               Chintala, Soumith",
  journal   = "Adv. Neural Inf. Process. Syst.",
  publisher = "proceedings.neurips.cc",
  volume    =  32,
  year      =  2019
}

@ARTICLE{NeuWS,
  title     = "{NeuWS}: Neural wavefront shaping for guidestar-free imaging
               through static and dynamic scattering media",
  author    = "Feng, Brandon Y and Guo, Haiyun and Xie, Mingyang and
               Boominathan, Vivek and Sharma, Manoj K and Veeraraghavan, Ashok
               and Metzler, Christopher A",
  journal   = "Sci. Adv.",
  publisher = "science.org",
  volume    =  9,
  number    =  26,
  month     =  jun,
  year      =  2023
}

@article{chen2025multilook,
  title={Multilook coherent imaging: Theoretical guarantees and algorithms},
  author={Chen, Xi and Jana, Soham and Metzler, Christopher A and Maleki, Arian and Jalali, Shirin},
  journal={arXiv preprint arXiv:2505.23594},
  year={2025}
}

@INPROCEEDINGS{Ziemann2024,
  title     = "Time-varying implicit neural representations for unsupervised
               speckle denoising in dynamic scenes",
  author    = "Ziemann, Matthew R and Joshi, Rushil R and Metzler, Christopher A",
  booktitle = "Unconventional Imaging, Sensing, and Adaptive Optics 2024",
  publisher = "SPIE",
  volume    =  13149,
  month     =  oct,
  year      =  2024
}

@ARTICLE{Bianco2018,
  title     = "Strategies for reducing speckle noise in digital holography",
  author    = "Bianco, Vittorio and Memmolo, Pasquale and Leo, Marco and
               Montresor, Silvio and Distante, Cosimo and Paturzo, Melania and
               Picart, Pascal and Javidi, Bahram and Ferraro, Pietro",
  journal   = "Light Sci. Appl.",
  publisher = "Springer Science and Business Media LLC",
  volume    =  7,
  number    =  1,
  pages     =  48,
  month     =  aug,
  year      =  2018,
  language  = "en"
}

@ARTICLE{Argenti2013,
  title     = "A tutorial on speckle reduction in synthetic aperture radar
               images",
  author    = "Argenti, Fabrizio and Lapini, Alessandro and Bianchi, Tiziano and
               Alparone, Luciano",
  journal   = "IEEE Geosci. Remote Sens. Mag.",
  publisher = "Institute of Electrical and Electronics Engineers (IEEE)",
  volume    =  1,
  number    =  3,
  pages     = "6--35",
  month     =  sep,
  year      =  2013
}

@ARTICLE{Ma2020,
  title     = "{SAR} image despeckling by noisy reference-based deep learning
               method",
  author    = "Ma, Xiaoshuang and Wang, Chen and Yin, Zhixiang and Wu, Penghai",
  journal   = "IEEE Trans. Geosci. Remote Sens.",
  publisher = "Institute of Electrical and Electronics Engineers (IEEE)",
  volume    =  58,
  number    =  12,
  pages     = "8807--8818",
  month     =  dec,
  year      =  2020
}

@ARTICLE{Lee1980,
  title     = "Digital image enhancement and noise filtering by use of local
               statistics",
  author    = "Lee, Jong-Sen",
  journal   = "IEEE Trans. Pattern Anal. Mach. Intell.",
  volume    = 2,
  number    = 2,
  pages     = "165--168",
  year      = 1980
}

@ARTICLE{Frost1982,
  title     = "A model for radar images and its application to adaptive digital
               filtering of multiplicative noise",
  author    = "Frost, Victor S and Stiles, Josephine Abbott and Shanmugan,
               K Sam and Holtzman, Julian C",
  journal   = "IEEE Trans. Pattern Anal. Mach. Intell.",
  volume    = 4,
  number    = 2,
  pages     = "157--166",
  year      = 1982
}

@ARTICLE{Parrilli2012,
  title     = "A nonlocal {SAR} image denoising algorithm based on {LLMMSE}
               wavelet shrinkage",
  author    = "Parrilli, Sara and Poderico, Mariana and Angelino, Cesario V
               and Verdoliva, Luisa",
  journal   = "IEEE Trans. Geosci. Remote Sens.",
  volume    = 50,
  number    = 2,
  pages     = "606--616",
  month     = feb,
  year      = 2012
}

@INPROCEEDINGS{Lehtinen2018,
  title     = "{Noise2Noise}: Learning image restoration without clean data",
  author    = "Lehtinen, Jaakko and Munkberg, Jacob and Hasselgren, Jon and
               Laine, Samuli and Karras, Tero and Aittala, Miika and Aila,
               Timo",
  booktitle = "Proceedings of the 35th International Conference on Machine
               Learning (ICML)",
  volume    = 80,
  pages     = "2965--2974",
  year      = 2018
}

@ARTICLE{Molini2022,
  title     = "{Speckle2Void}: Deep self-supervised {SAR} despeckling with
               blind-spot convolutional neural networks",
  author    = "Bordone Molini, Andrea and Valsesia, Diego and Fracastoro,
               Giulia and Magli, Enrico",
  journal   = "IEEE Trans. Geosci. Remote Sens.",
  volume    = 60,
  pages     = "1--17",
  year      = 2022
}

@INPROCEEDINGS{Tancik2020,
  title     = "Fourier features let networks learn high frequency functions in
               low dimensional domains",
  author    = "Tancik, Matthew and Srinivasan, Pratul P and Mildenhall, Ben
               and Fridovich-Keil, Sara and Raghavan, Nithin and Singhal,
               Utkarsh and Ramamoorthi, Ravi and Barron, Jonathan T and Ng,
               Ren",
  booktitle = "Advances in Neural Information Processing Systems (NeurIPS)",
  volume    = 33,
  pages     = "7537--7547",
  year      = 2020
}

@INPROCEEDINGS{wire,
  title     = "{WIRE}: Wavelet implicit neural representations",
  author    = "Saragadam, Vishwanath and LeJeune, Daniel and Tan, Jasper and
               Balakrishnan, Guha and Veeraraghavan, Ashok and Baraniuk, Richard
               G",
  booktitle = "2023 IEEE/CVF Conference on Computer Vision and Pattern
               Recognition (CVPR)",
  publisher = "IEEE",
  month     =  jun,
  year      =  2023
}

@article{hutchinson,
author = {M.F. Hutchinson},
title = {A Stochastic Estimator of the Trace of the Influence Matrix for Laplacian Smoothing Splines},
journal = {Communications in Statistics - Simulation and Computation},
volume = {18},
number = {3},
pages = {1059--1076},
year = {1989},
publisher = {Taylor \& Francis},
}

@ARTICLE{cg,
  title     = "Methods of conjugate gradients for solving linear systems",
  author    = "Hestenes, M R and Stiefel, E",
  journal   = "Journal of research of the National Bureau of Standards",
  publisher = "books.google.com",
  volume    =  49,
  number    =  6,
  pages     = "409--436",
  year      =  1952
}

@ARTICLE{slq,
  title     = "Fast estimation of $tr(f(A))$ via stochastic lanczos quadrature",
  author    = "Ubaru, Shashanka and Chen, Jie and Saad, Yousef",
  journal   = "SIAM J. Matrix Anal. Appl.",
  publisher = "Society for Industrial \& Applied Mathematics (SIAM)",
  volume    =  38,
  number    =  4,
  pages     = "1075--1099",
  month     =  jan,
  year      =  2017,
  language  = "en"
}

@article{cao2024neural,
  title={Neural space--time model for dynamic multi-shot imaging},
  author={Cao, Ruiming and Divekar, Nikita S and Nunez, James K and Upadhyayula, Srigokul and Waller, Laura},
  journal={Nature Methods},
  volume={21},
  number={12},
  pages={2336--2341},
  year={2024},
  publisher={Nature Publishing Group US New York}
}

@inproceedings{scalable_gauss_processes,
 author = {Dong, Kun and Eriksson, David and Nickisch, Hannes and Bindel, David and Wilson, Andrew G},
 booktitle = {Advances in Neural Information Processing Systems},
 editor = {I. Guyon and U. Von Luxburg and S. Bengio and H. Wallach and R. Fergus and S. Vishwanathan and R. Garnett},
 pages = {},
 publisher = {Curran Associates, Inc.},
 title = {Scalable Log Determinants for Gaussian Process Kernel Learning},
 url = {https://proceedings.neurips.cc/paper_files/paper/2017/file/976abf49974d4686f87192efa0513ae0-Paper.pdf},
 volume = {30},
 year = {2017}
}

@ARTICLE{mulog,
  title     = "{MuLoG}, or how to apply Gaussian denoisers to multi-channel
               {SAR} speckle reduction?",
  author    = "Deledalle, Charles-Alban and Denis, Loic and Tabti, Sonia and
               Tupin, Florence",
  journal   = "IEEE Trans. Image Process.",
  publisher = "IEEE",
  volume    =  26,
  number    =  9,
  pages     = "4389--4403",
  month     =  sep,
  year      =  2017,
  language  = "en"
}

@ARTICLE{drunet,
  title     = "Plug-and-play image restoration with deep denoiser prior",
  author    = "Zhang, Kai and Li, Yawei and Zuo, Wangmeng and Zhang, Lei and Van
               Gool, Luc and Timofte, Radu",
  journal   = "IEEE Trans. Pattern Anal. Mach. Intell.",
  publisher = "Institute of Electrical and Electronics Engineers (IEEE)",
  volume    =  44,
  number    =  10,
  pages     = "6360--6376",
  month     =  oct,
  year      =  2022,
  language  = "en"
}

@ARTICLE{DAVIS-2017,
  title         = "The 2017 {DAVIS} challenge on Video Object Segmentation",
  author        = "Pont-Tuset, Jordi and Perazzi, Federico and Caelles, Sergi
                   and Arbeláez, Pablo and Sorkine-Hornung, Alex and Van Gool,
                   Luc",
  journal       = "arXiv preprint arXiv:1704.00675 ",
  month         =  apr,
  year          =  2017,
  archivePrefix = "arXiv",
  primaryClass  = "cs.CV"
}

@inproceedings{fastdvdnet,
  title={Fastdvdnet: Towards real-time deep video denoising without flow estimation},
  author={Tassano, Matias and Delon, Julie and Veit, Thomas},
  booktitle={Proceedings of the IEEE/CVF conference on computer vision and pattern recognition},
  pages={1354--1363},
  year={2020}
}

@ARTICLE{DVP,
  title     = "Deep Video Prior for video consistency and propagation",
  author    = "Lei, Chenyang and Xing, Yazhou and Ouyang, Hao and Chen, Qifeng",
  journal   = "IEEE Trans. Pattern Anal. Mach. Intell.",
  publisher = "Institute of Electrical and Electronics Engineers (IEEE)",
  volume    =  45,
  number    =  1,
  pages     = "356--371",
  month     =  jan,
  year      =  2023,
  language  = "en"
}

@article{spencer2017,
	Author = {Spencer, Mark F and Raynor, Robert A and Banet, Matthias T and Marker, Dan K},
	Journal = {Optical Engineering},
	Number = {3},
	Pages = {031213--031213},
	Publisher = {International Society for Optics and Photonics},
	Title = {Deep-turbulence wavefront sensing using digital-holographic detection in the off-axis image plane recording geometry},
	Volume = {56},
	Year = {2017}}

@article{thurman2019,
	Author = {Thurman, Samuel T},
	Journal = {JOSA A},
	Number = {12},
	Pages = {D47--D61},
	Publisher = {Optical Society of America},
	Title = {Phase-error correction in digital holography using single-shot data},
	Volume = {36},
	Year = {2019}}

@ARTICLE{strred,
  title     = "Video quality assessment by reduced reference spatio-temporal
               entropic differencing",
  author    = "Soundararajan, Rajiv and Bovik, Alan C",
  journal   = "IEEE Trans. Circuits Syst. Video Technol.",
  publisher = "Institute of Electrical and Electronics Engineers (IEEE)",
  volume    =  23,
  number    =  4,
  pages     = "684--694",
  month     =  apr,
  year      =  2013
}

@ARTICLE{ssim,
  title     = "Image quality assessment: from error visibility to structural
               similarity",
  author    = "Wang, Zhou and Bovik, Alan Conrad and Sheikh, Hamid Rahim and
               Simoncelli, Eero P",
  journal   = "IEEE Trans. Image Process.",
  publisher = "Institute of Electrical and Electronics Engineers (IEEE)",
  volume    =  13,
  number    =  4,
  pages     = "600--612",
  month     =  apr,
  year      =  2004,
  language  = "en"
}

@ARTICLE{swin-unet,
  title     = "Speckle denoising based on Swin-{UNet} in digital holographic
               interferometry",
  author    = "Chen, Jie and Liao, Houzhang and Kong, Yong and Zhang, Dawei and
               Zhuang, Songlin",
  journal   = "Opt. Express",
  publisher = "Optica Publishing Group",
  volume    =  32,
  number    =  19,
  pages     = "33465--33482",
  month     =  sep,
  year      =  2024,
  language  = "en"
}

@ARTICLE{Allen2025CLAMP,
  title     = "{CLAMP}: Majorized plug-and-play for coherent {3D} lidar imaging",
  author    = "Allen, Tony G and Rabb, David J and Buzzard, Gregery T and
               Bouman, Charles A",
  journal   = "IEEE Trans. Comput. Imaging",
  publisher = "Institute of Electrical and Electronics Engineers (IEEE)",
  volume    =  11,
  pages     = "506--519",
  year      =  2025,
  language  = "en"
}

@inproceedings{DERF-HD,
  author    = {Arias, Pablo and Facciolo, Gabriele and Morel, Jean-Michel},
  title     = {A Comparison of Patch-Based Models in Video Denoising},
  booktitle = {2018 IEEE 13th Image, Video, and Multidimensional Signal Processing Workshop (IVMSP)},
  year      = {2018},
  pages     = {1--5},
  doi       = {10.1109/IVMSPW.2018.8448573}
}

@misc{derf,
  author       = {{Xiph.org Foundation}},
  title        = {{Xiph.org Video Test Media [derf's collection]}},
  howpublished = {\url{https://media.xiph.org/video/derf/}},
}

@ARTICLE{Chen2026,
  title         = "Maximum likelihood reconstruction for multi-look digital
                   holography with {M}arkov-modeled speckle correlation",
  author        = "Chen, Xi and Maleki, Arian and Jalali, Shirin",
  journal       = "arXiv preprint arXiv:2604.20154",
  month         =  apr,
  year          =  2026,
  archivePrefix = "arXiv",
  primaryClass  = "eess.IV"
}

@ARTICLE{golay,
  title     = "A golay metalens for long-range, large aperture, thermal imaging
               via sparse aperture computational imaging",
  author    = "Wang, Jing and Wirth-Singh, Anna and Saragadam, Vishwanath and
               Johnson, Rose and Majumdar, Arka and Veeraraghavan, Ashok",
  journal   = "Nat. Commun.",
  publisher = "Springer Science and Business Media LLC",
  volume    =  16,
  number    =  1,
  pages     =  10281,
  month     =  nov,
  year      =  2025,
  language  = "en"
}

@ARTICLE{Dalsasso2022,
  title     = "As if by magic: Self-supervised training of deep despeckling
               networks with {MERLIN}",
  author    = "Dalsasso, Emanuele and Denis, Loic and Tupin, Florence",
  journal   = "IEEE Trans. Geosci. Remote Sens.",
  publisher = "Institute of Electrical and Electronics Engineers (IEEE)",
  volume    =  60,
  pages     = "1--13",
  year      =  2022
}

\end{document}


\begin{abstract}
\end{abstract}

\maketitle

\section{Aperture-aware likelihood derivation}
\label{sec:supp_likelihood}

This section provides the full likelihood derivation summarized in the main text. For a discrete spatial grid with $N$ pixels and $T$ temporal measurements, let $r_t\in\mathbb{R}_+^N$ denote the nonnegative scene reflectance at time $t$. Under the fully developed speckle model, the object-plane field is
\begin{equation}
    s_t = r_t^{1/2}\circ g_t,
    \qquad
    g_t\sim\mathcal{CN}(0,I),
\end{equation}
where $\circ$ denotes elementwise multiplication. Coherent propagation through the aperture is modeled by
\begin{equation}
    A = F^H\mathcal{D}(p)F,
\end{equation}
with $F$ the unitary two-dimensional DFT and $p$ the sampled pupil transmission. The measured complex field is
\begin{equation}
    \tilde{y}_t = A(r_t^{1/2}\circ g_t) + n_t,
    \qquad
    n_t\sim\mathcal{CN}(0,\sigma^2 I).
\end{equation}
Conditional on $r_t$, the speckled object-plane field is complex Gaussian with zero mean and diagonal covariance $\mathcal{D}(r_t)$. Since $A$ is linear and $n_t$ is independent of $g_t$, the measurement is also complex Gaussian with covariance
\begin{equation}
    \Sigma_t
    = \mathbb{E}\left[\tilde{y}_t\tilde{y}_t^H\mid r_t\right]
    = A\mathcal{D}(r_t)A^H + \sigma^2 I.
\end{equation}
Thus,
\begin{equation}
    \tilde{y}_{t} \mid r_{t}
    \sim \mathcal{CN} \bigl( 0, \Sigma_{t} \bigr),
\end{equation}
with density
\begin{equation}
    p(\tilde{y}_{t} \mid r_{t})
    = \frac{1}{\pi^{N}\det(\Sigma_{t})}
    \exp\!\left(-\tilde{y}_{t}^H \Sigma_{t}^{-1} \tilde{y}_{t}\right).
\end{equation}
The per-frame negative log-likelihood, omitting additive constants, is therefore
\begin{equation}
    \mathcal{L}_{t}(r_t)
    = \log \det \bigl(\Sigma_{t}\bigr)
    + \tilde{y}_{t}^H \Sigma_{t}^{-1} \tilde{y}_{t}.
\end{equation}
Assuming independent speckle and additive-noise realizations across time yields the sequence-level objective used in the main text,
\begin{equation}
    \mathcal{L}_{\mathrm{MLE}}
    = \sum_{t}
    \left[
        \log \det \bigl(\Sigma_t\bigr)
        + \tilde{y}_{t}^H\Sigma_t^{-1}\tilde{y}_{t}
    \right].
\end{equation}

\section{Matrix-free likelihood implementation}
\label{sec:supp_matrix_free}

The proposed MLE objective requires evaluating, for each temporal measurement, a log determinant and an inverse quadratic form involving the measurement covariance. For a single frame, we write the covariance and per-frame negative log-likelihood as
\begin{equation}
    \Sigma_t = A\mathcal{D}(r_t)A^H + \sigma^2 I,
    \label{eq:supp_covariance}
\end{equation}
\begin{equation}
    \mathcal{L}_{t}(r_t)
    = \log\det\bigl(\Sigma_t\bigr)
    + \tilde{y}_{t}^H \Sigma_{t}^{-1} \tilde{y}_{t}.
    \label{eq:supp_frame_loss}
\end{equation}
A direct implementation of this objective forms $\Sigma_t$ explicitly and computes both its log-determinant and its inverse for every frame and every iteration of INR fitting. This yields $\mathcal{O}(N^3)$ time and $\mathcal{O}(N^2)$ memory per frame, which leads to very long reconstruction times for even modest spatial resolutions and quickly becomes intractable for realistic image sizes.

In this work, we accelerate evaluation of the MLE loss by replacing the dense linear algebra with a matrix-free implementation that dramatically reduces both memory usage and runtime. Concretely, we exploit three ideas: \emph{(i)} FFT-based application of the forward model, \emph{(ii)} stochastic Lanczos quadrature (SLQ) for the log-determinant, and \emph{(iii)} conjugate gradients (CG) for the quadratic form.

\subsection{FFT-based application of the covariance}

The core improvement to our objective calculation is to avoid explicitly forming $\Sigma_t$ by leveraging 2D fast Fourier transforms (FFTs) in place of the discrete Fourier transform matrices. Under the zero-turbulence assumption used in this work, the measurement operator is written as
\begin{equation}
    A = F^H\mathcal{D}(p)F,
    \label{eq:supp_A_definition}
\end{equation}
where $F$ is the unitary two-dimensional discrete Fourier transform (DFT) and $p$ is the real-valued entrance-pupil transmission function. Substituting this into \eqref{eq:supp_covariance} yields
\begin{equation}
    \Sigma_t = F^H\mathcal{D}(p)F\mathcal{D}(r_t)F^H\mathcal{D}(p)F + \sigma^2 I.
    \label{eq:supp_covariance_fft}
\end{equation}
Because $N$ is the number of pixels in a single vectorized image frame of height $H$ and width $W$, $\Sigma_t$ is a dense $N\times N$ covariance matrix that grows impractically large for realistic image resolutions.

Rather than explicitly forming $\Sigma_t$, we implement the matrix--vector product $\Sigma_t\xi$ by applying the DFT operators using 2D FFTs. Letting $\mathcal{F}$ and $\mathcal{F}^{-1}$ denote the 2D FFT and its inverse, we have
\begin{equation}
    \Sigma_t\xi
    =
    \mathcal{F}^{-1}\!\left(
    P \odot \mathcal{F}\!\left(
    R_t \odot \mathcal{F}^{-1}\!\left(P\odot \mathcal{F}(\xi)\right)
    \right)
    \right)
    +\sigma^2 \xi,
    \label{eq:supp_covariance_matvec}
\end{equation}
where $\xi\in\mathbb{C}^{H\times W}$ denotes an arbitrary input reshaped onto the 2D measurement grid, $R_t \in \mathbb{R}^{H\times W}$ is the 2D reflectance matrix, and $P\in \mathbb{R}^{H\times W}$ is the 2D aperture mask.

Each application of $\Sigma_t$ therefore requires two forward FFTs, two inverse FFTs, and a small number of element-wise operations, yielding $\mathcal{O}(N \log N)$ complexity per matrix--vector product. Related likelihood-based coherent-reconstruction methods have accelerated repeated covariance inversions using Newton--Schulz iterations within projected-gradient optimization~\cite{chen2025multilook}. Our formulation instead avoids explicit inversion and factorization by using the FFT-based covariance matrix--vector product to evaluate both terms in the MLE loss in \eqref{eq:supp_frame_loss}, with stochastic Lanczos quadrature for the log-determinant and conjugate gradients for the quadratic form.

\subsection{Log-determinant via stochastic Lanczos quadrature}

The first term in \eqref{eq:supp_frame_loss} is the log-determinant of the covariance, $\log\det(\Sigma_t)$. To avoid forming $\Sigma_t$ explicitly, we estimate this term using stochastic Lanczos quadrature (SLQ)~\cite{slq,scalable_gauss_processes}. Using the trace--log identity,
\begin{equation}
    \log\det(\Sigma_t) = \operatorname{tr}\!\left(\log \Sigma_t\right),
\end{equation}
we approximate the trace with $S$ Hutchinson probes~\cite{hutchinson}:
\begin{equation}
    \operatorname{tr}\!\left(\log \Sigma_t\right)
    \approx
    \frac{1}{S}\sum_{s=1}^S 
    z_s^H \log(\Sigma_t) z_s,
\end{equation}
where $z_s$ are random probe vectors satisfying $\mathbb{E}[z_s z_s^H]=I$. Each quadratic form $z_s^H \log(\Sigma_t) z_s$ is then approximated by running $K$ steps of Hermitian Lanczos using only matrix--vector products with $\Sigma_t$:
\begin{equation}
    z_s^H \log(\Sigma_t) z_s
    \approx
    \|z_s\|_2^2 e_1^T \log(T_s)e_1,
\end{equation}
where $e_1$ is the first canonical basis vector and $T_s\in\mathbb{R}^{K\times K}$ is the tridiagonal matrix produced by $K$ steps of Lanczos corresponding to probe $z_s$. Since each Lanczos step requires one FFT-based covariance matrix--vector product, the log-determinant estimate costs $\mathcal{O}(S K N \log N)$ per frame.

\subsection{Quadratic form via conjugate gradients}

The second term in \eqref{eq:supp_frame_loss} is the data-fit quadratic form $\tilde{y}_{t}^H \Sigma_{t}^{-1} \tilde{y}_{t}$. Evaluating this term does not require explicitly constructing $\Sigma_{t}^{-1}$; it only requires solving a linear system involving $\Sigma_t$. Specifically, we define
\begin{equation}
    x_t = \Sigma_t^{-1}\tilde{y}_t
    \quad \Longleftrightarrow \quad
    \Sigma_tx_t = \tilde{y}_t.
\end{equation}
Since $\Sigma_t=A\mathcal{D}(r_t)A^H+\sigma^2 I$ is Hermitian positive definite for $\sigma^2>0$, this linear system can be solved efficiently using conjugate gradients (CG)~\cite{cg}. Once $x_t$ is obtained, the quadratic term is evaluated as
\begin{equation}
    \tilde{y}_{t}^H \Sigma_{t}^{-1} \tilde{y}_{t}
    =
    \tilde{y}_{t}^H x_t.
\end{equation}
Each CG iteration requires only a matrix--vector product with $\Sigma_t$, which we compute using the FFT-based covariance application described above. Thus, the quadratic term is evaluated without forming $\Sigma_t$ or its inverse. If $G$ CG iterations are used, the cost scales as $\mathcal{O}(G\,N\log N)$ per frame.

\subsection{Resulting complexity}

Putting these components together, our accelerated MLE loss approximates the maximum likelihood objective by replacing dense matrix formation, inversion, and factorization with FFT-based matrix--vector products, using SLQ for the log-determinant and CG for the quadratic term. We implement this approximation in PyTorch~\cite{pytorch}, allowing gradients to be propagated with respect to the INR parameters during training. This yields an implementation whose per-frame cost scales as $\mathcal{O}((SK + G)N\log N)$, where $S$ is the number of Hutchinson probes, $K$ is the number of Lanczos iterations, and $G$ is the number of CG iterations. In practice, we use 8 Hutchinson probes, 20 Lanczos iterations, and 200 maximum CG iterations. Additionally, the memory footprint remains linear in $N$, enabling practical optimization of the INR on higher-resolution sequences and larger temporal windows.

\subsection{Additional runtime context and approximation accuracy}
\label{sec:supp_scaling}

The main text reports the runtime scaling of the proposed matrix-free MLE loss relative to an explicit covariance implementation. Here we provide additional runtime context and evaluate the approximation error introduced by replacing dense matrix operations with FFT-based matrix--vector products, stochastic Lanczos quadrature, and conjugate gradients.

For context, the full likelihood-based INR used in the main experiments requires an average of 85 seconds per frame at $256\times256$ resolution on an NVIDIA RTX Pro Blackwell GPU. This is substantially slower than feed-forward supervised baselines such as DRUNet and FastDVDNet, which require 9 and 18 hours of offline training respectively on four NVIDIA A100 GPUs, but run at approximately 3 milliseconds per frame at test time. The same WIRE architecture trained with a simpler MSE loss requires 53 seconds per frame, indicating that the runtime bottleneck is not only fitting the INR itself, but also repeatedly evaluating the MLE objective through the SLQ log-determinant estimate and CG quadratic form.

\begin{figure}[t]
    \centering
    \includegraphics[width=0.85\linewidth]{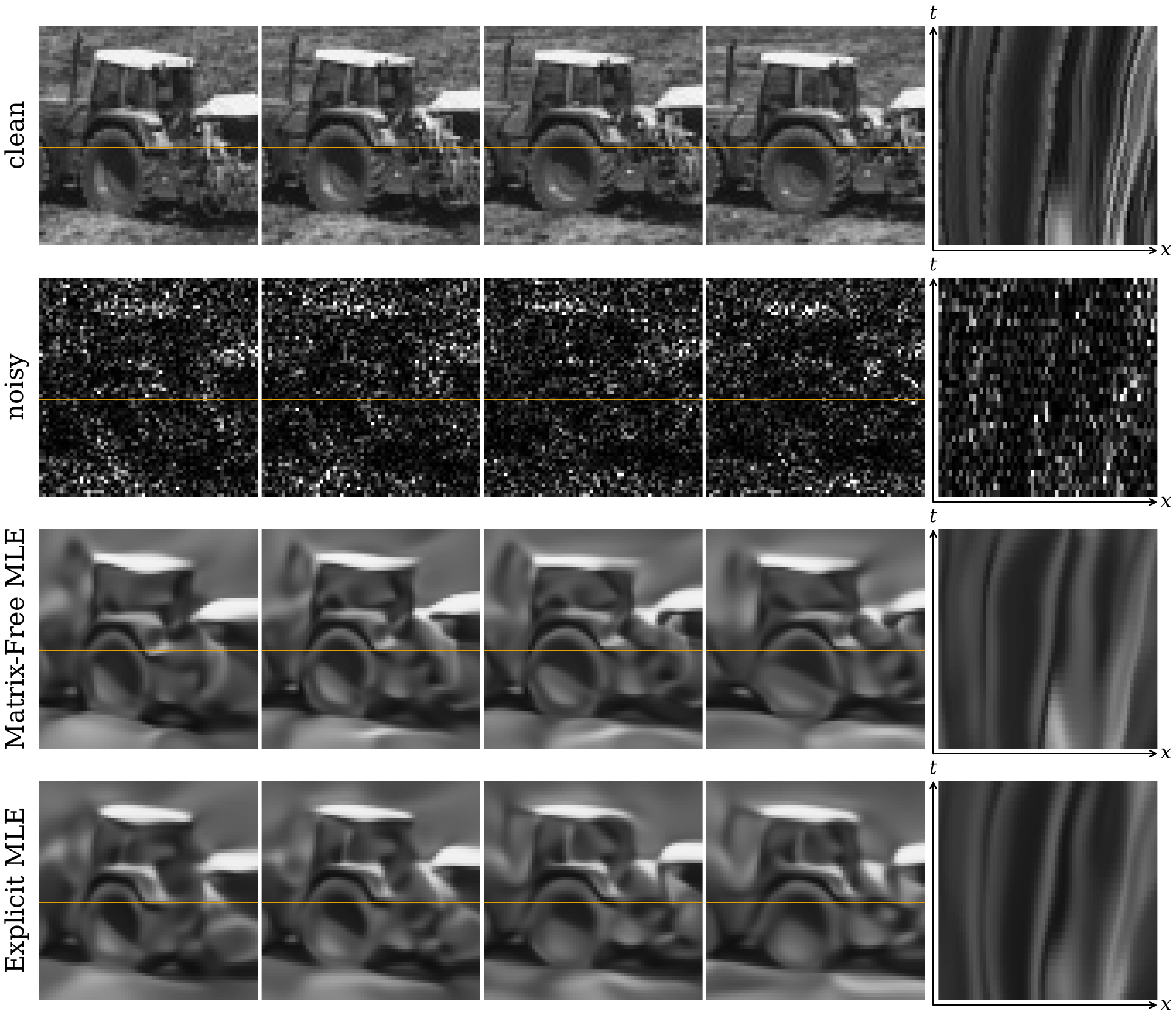}
    \caption{\textbf{Effect of MLE approximation on denoising.} Comparison of our matrix-free MLE loss approximation and explicit MLE formulation on a $64\times64$ resolution DERF-HD \texttt{tractor} sequence. Our approximation yields comparable results with significantly better scaling.}
    \label{fig:supp_mle_ablation}
\end{figure}

The MLE loss approximation introduces only a modest reduction in spatial reconstruction quality, as seen in Fig.~\ref{fig:supp_mle_ablation}. When compared to the explicit implementation on the DERF-HD \texttt{tractor} sequence at $64\times64$ resolution, our approximation reduces denoised PSNR from 22.4 dB to 22.0 dB and SSIM from 0.71 to 0.69. ST-RRED increases more substantially, from 13486 to 22945; however, this numerical increase does not correspond to a clear visual degradation in temporal consistency. As shown in the $x$--$t$ plots in Fig.~\ref{fig:supp_mle_ablation}, both methods retain smooth lines and consistent trajectories. We attribute this discrepancy in part to the low spatial resolution of the comparison, where small frame-wise differences can produce large changes in the metric. The low resolution of this example is necessary for practical runtime of the explicit method, which would be impractical for repeated INR optimization at $128\times128$. This low-resolution setting also makes denoising especially challenging because image features occupy only a few pixels and are therefore comparable to the speckle grain size.

Overall, the matrix-free formulation trades minor approximation error for a substantial improvement in scalability, making the proposed MLE objective practical for realistic coherent imaging data. Though we retain constant solver parameters for this work, future work could explore coarse-to-fine solver tuning and other methods to reduce approximation error.

\section{Baseline and evaluation details}

This section provides the implementation details for the classical, untrained neural, and supervised baselines summarized in the main text. All neural baselines are implemented in PyTorch~\cite{pytorch}.

\paragraph{H-BM3D.}
We utilize homomorphic BM3D as a traditional single-frame baseline. BM3D was originally developed for additive white Gaussian noise~\cite{BM3D}, so we first apply a logarithmic transform to the measured intensity in order to approximately convert multiplicative speckle into additive noise. We use the standard log-speckle variance for single-look speckle, denoise the log-intensity image with BM3D, and then invert the transform. This provides a strong traditional baseline that requires no training data or temporal information, while following the common homomorphic strategy used for speckle denoising~\cite{mulog}.

\paragraph{DVP.}
We compare against Deep Video Prior (DVP), an unsupervised multi-frame neural denoising method that fits a convolutional network directly to the corrupted video sequence~\cite{DVP}. DVP receives the complex field measurements as input and is optimized using an intensity-domain MSE loss. Like our method, DVP does not require an external training dataset and instead relies on the implicit regularization of the network architecture. We fit DVP to a 32-frame windowed sequence as a direct comparison to our method. It has 7.7 million trainable parameters.

\paragraph{DRUNet.}
We include DRUNet as a supervised single-frame image-denoising baseline~\cite{drunet}. DRUNet operates on intensity images and is trained for 500 epochs using random crops from the DIV2K dataset~\cite{div2k}, resized to $256\times256$ pixels. Training data are generated in-domain by applying the simulated coherent imaging forward model with $\sigma=0.02$. The standard DRUNet model is trained using a simplified pixelwise speckle model without an aperture, corresponding to the limiting case of uncorrelated speckle. We also evaluate aperture-specific DRUNet-A variants, which are retrained separately for each aperture to test the benefit of matching the training distribution to the spatial speckle correlations. Both models have 33 million trainable parameters.

\paragraph{FastDVDNet.}
We compare against FastDVDNet as a supervised video-denoising baseline~\cite{fastdvdnet}. FastDVDNet operates on intensity videos and is trained for 500 epochs on DAVIS-2017 videos~\cite{DAVIS-2017}, using per-epoch random crops resized to $256\times256$ pixels. As with DRUNet, the training data are generated by applying the simulated coherent imaging forward model with $\sigma=0.02$. FastDVDNet takes five neighboring frames as input and predicts the denoised center frame, allowing it to exploit local temporal context. The standard FastDVDNet model is trained using the simplified pixelwise speckle model, while FastDVDNet-A variants are retrained separately for each aperture. Both variants have 2.5 million trainable parameters.

\paragraph{Metrics and stopping criteria.}
We quantify simulated reconstruction quality using PSNR, SSIM~\cite{ssim}, and ST-RRED~\cite{strred} relative to clean ground-truth sequences. PSNR and SSIM measure frame-wise spatial fidelity, while ST-RRED measures spatiotemporal perceptual distortion and is sensitive to temporal flickering artifacts. For DVP, we use oracle early stopping to report its best attainable reconstruction quality, which typically occurs within $10$--$20$ optimization iterations. In contrast, our method uses the holdout early-stopping criterion described in the main text Section 2.3. Specifically, we withhold two frames (indices 12 and 28) from each 32-frame sequence and evaluate their MLE loss once per optimization iteration. When a holdout loss minimum is found, model weights are checkpointed, and training is stopped when the holdout loss minimum does not decrease after a patience threshold. Held-out frames are included in PSNR, SSIM, and ST-RRED evaluations. 

For the simulated data, we use a patience threshold of 150 iterations. We found the laboratory data required greater patience, as the resolution bar data caused premature stopping with 150 iteration patience due to its sparse, predominantly low-frequency features. For our presented laboratory results, we use an early stopping patience of 750 iterations, and smooth the holdout loss over a 15 iteration window to reduce the effects of noise. On average, our early stopping criterion stops optimization after approximately 400 iterations on the simulated data and 2400 iterations on the laboratory data. The laboratory data requires more iterations due to slow fitting of the high frequency areas of the bar targets. Adjustment of learning rate may improve fitting times, but we chose to keep the model hyperparameters fixed for this work.

\section{Simulated data generation details}
\label{sec:supp_sim_details}

The simulated experiments use both a controlled translating-image sequence and the DERF-HD video-denoising benchmark~\cite{DERF-HD}. For the toy sequence, we translate a $256\times256$ crop diagonally across the \texttt{cameraman} image over 32 frames. For DERF-HD, we use seven 100-frame, $960\times540$ grayscale sequences derived from the Xiph.Org DERF video test media collection~\cite{derf}. We use the first 32 frames of each video, center-crop each frame to a square aspect ratio, and resize it to $256\times256$ pixels.

For all simulated sequences, we generate complex-valued speckled measurements using the coherent forward model in Eq.~(2) of the main text with additive Gaussian noise level $\sigma=0.02$. To evaluate robustness to aperture-dependent spatial correlation, we simulate three aperture variants at fixed sensor-plane sampling: a large circular aperture with diameter equal to the full computational grid width (256 pixels), a small circular aperture with diameter 128 pixels, and a sparse Golay aperture~\cite{golay}.

The large-aperture case produces the shortest speckle correlation length, whereas the small aperture increases the spatial correlation, producing larger speckle grains and a more challenging denoising problem. The Golay aperture is based on the distributed Golay 6+1 metalens geometry of Wang et al.~\cite{golay} and consists of seven equal-diameter circular sub-apertures: one centered on the optical axis and six placed at asymmetric off-axis locations. We model the pupil as
\begin{equation}
P(\mathbf{r})
=
\sum_{i=0}^{6}
\operatorname{circ}
\left(
\frac{2\lVert \mathbf{r}-\mathbf{r}_i\rVert}{d}
\right),
\end{equation}
where $d$ is the sub-aperture diameter and $\mathbf{r}_i$ denotes the center of the $i$th sub-aperture. We uniformly scale the published Golay 6+1 geometry and sub-aperture dimensions to the computational pupil. Because the Golay pupil is sparse, disconnected, and asymmetric, it produces highly nonuniform spatial-frequency support, pronounced point-spread-function sidelobes, and anisotropic speckle correlations. It therefore provides a challenging test of adaptation to an irregular aperture.

\begin{figure}[h]
    \centering
    \includegraphics[width=\linewidth]{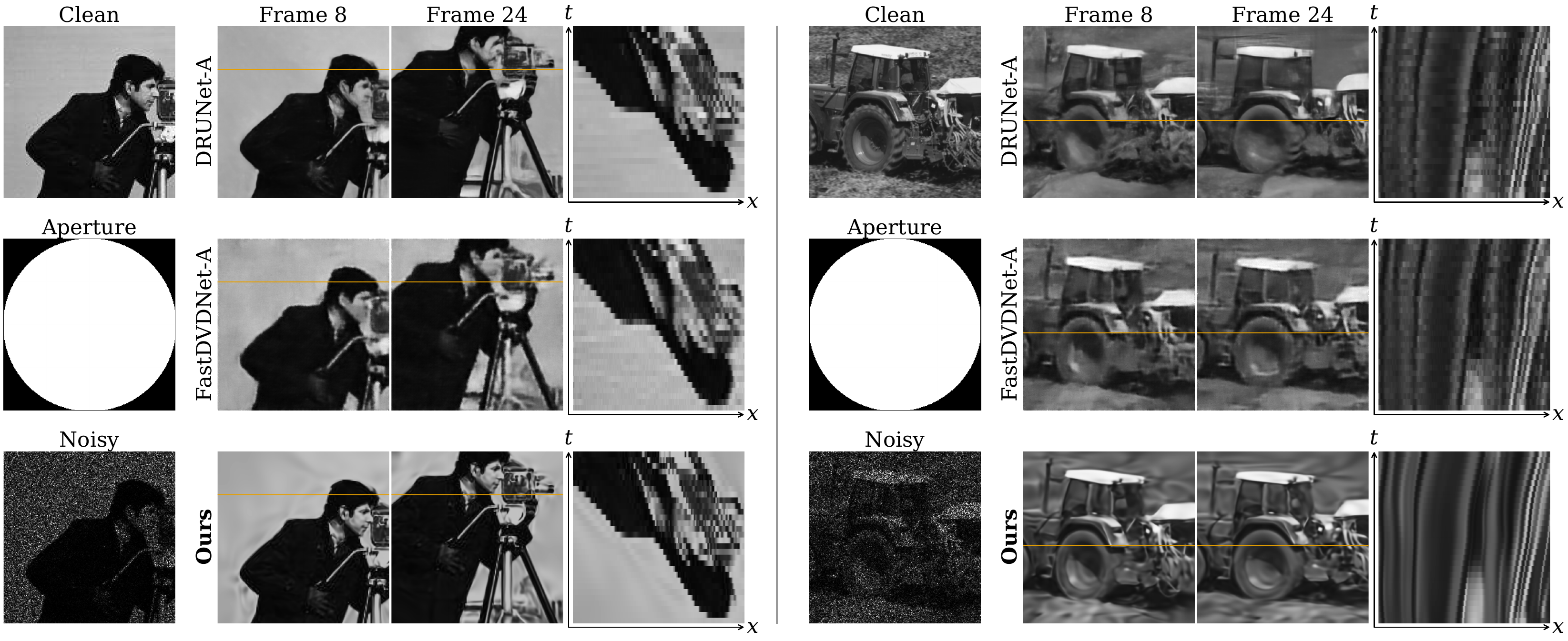}
    \caption{\textbf{Improved denoising on simulated data.} Simulated large-aperture speckle denoising on translating \texttt{cameraman} (left) and DERF-HD \texttt{tractor} (right) sequences. Our method recovers greater, temporally consistent detail.}
    \label{fig:sim_1A}
\end{figure}

\section{Additional simulated results}
\label{sec:supp_sim_results}

Our main text includes visualizations for the small and Golay apertures. For completeness, we include large aperture comparisons in Fig.~\ref{fig:sim_1A}. Denoising performance is clearer in video format available in Visualizations 8--9. These results show our method recovers greater detail in the face and camera of the \texttt{cameraman} sequence, as well as finer details in the landscape and tubing of the DERF-HD \texttt{tractor} sequence. Our method retains stable features across the entire denoised sequence, while other methods exhibit substantial frame-to-frame variation.

\section{Laboratory data collection details}
\label{sec:supp_lab_details}

To validate our approach with real data, we used the experimental digital holography (DH) setup shown in the main text to collect speckled video of moving objects. This setup has an off-axis image-plane geometry~\cite{spencer2017}. For our master oscillator (MO), we used a 2 W Coherent Verdi G single-frequency 532 nm source with an optical isolator to protect the source from back reflections. A half-wave plate (HWP) and polarizing beam splitter (PBS) created the signal and reference paths. To control the relative strength of each path, we added an additional HWP and PBS, which allowed us to further split the beam and send a tunable amount of light into a beam dump. We also added a variable neutral density (VND) filter to the reference path for fine tuning. In the signal path, we used a beam expander and steering mirror to flood illuminate our objects, located 2.6 m away. A 1 inch, 500 mm lens imaged the light reflected off the object onto our Phantom T3610 camera, which had a pixel pitch of 18 $\mu$m.

\begin{figure}[tbh]
    \centering
    \includegraphics[width=0.5\linewidth]{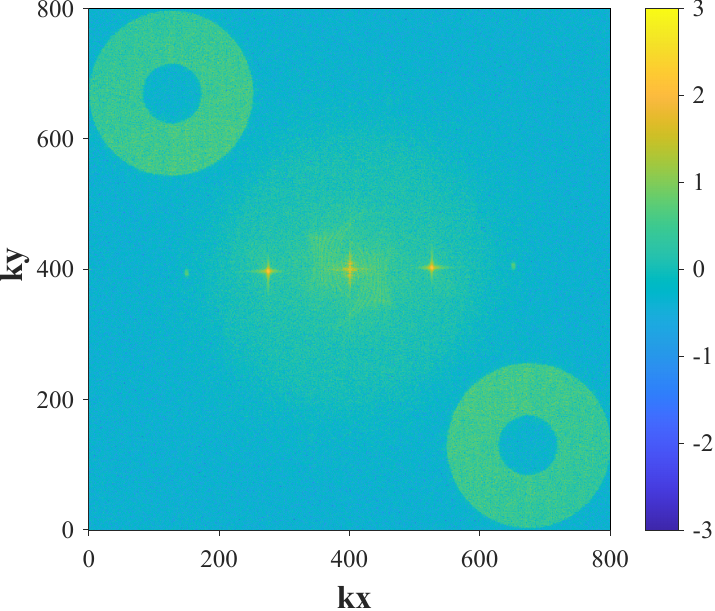}
    \caption{\textbf{Laboratory measured annular aperture support.} Fourier-domain magnitude (log scale) of the laboratory off-axis holograms from the modified digital holography setup with an inserted telescope. The system's annular aperture can be seen in the upper left and lower right of the image and is used to inform the likelihood model for the annular-aperture denoising results.}
    \label{fig:supp_lab_annular_kspace}
\end{figure}

To obtain a digital hologram, we mixed the light reflected off the object with light from our reference path, which passed through a 100 m single-mode fiber optic cable. We injected the reference light just after our imaging lens using a plate beam splitter. To control the sampling quotient of our DH system~\cite{spencer2017}, $q$, we used a variable iris on our imaging lens, which we adjusted until we obtained $q \approx 3$. This sampling quotient describes the relationship between detector sampling and the diffraction-limited spot size; for $q>1$, each speckle grain spans multiple detector samples, producing spatially correlated speckle in the measured field. To simulate motion, we placed the object on a single-axis translation stage and manually moved it while recording. For the annular-aperture experiment, we introduced a reflecting mode telescope into the beam path with measured $k$-space aperture support visible in Fig.~\ref{fig:supp_lab_annular_kspace}.

During each experiment, we collected 2,163 12-bit, $800 \times 800$ frames at 200 Hz with an exposure time of 50 $\mu$s. Prior to denoising, we temporally downsampled the video to 40 Hz by retaining every fifth frame, yielding larger inter-frame object displacements and providing a more challenging dynamic denoising setting. We spatially demodulated each digital hologram to obtain a $256 \times 256$-pixel, complex-valued field in the aperture plane of our sensor. For details about the demodulation process used for the off-axis IPRG, we direct the reader to Ref.~\cite{Pellizzari2017}. Finally, we used the image-sharpening method from Ref.~\cite{thurman2019} to estimate and digitally correct aberrations caused by static phase errors in our system.

\section{Additional laboratory results}
\label{sec:supp_lab_results}

\begin{figure}[h]
    \centering
    \includegraphics[width=0.9\linewidth]{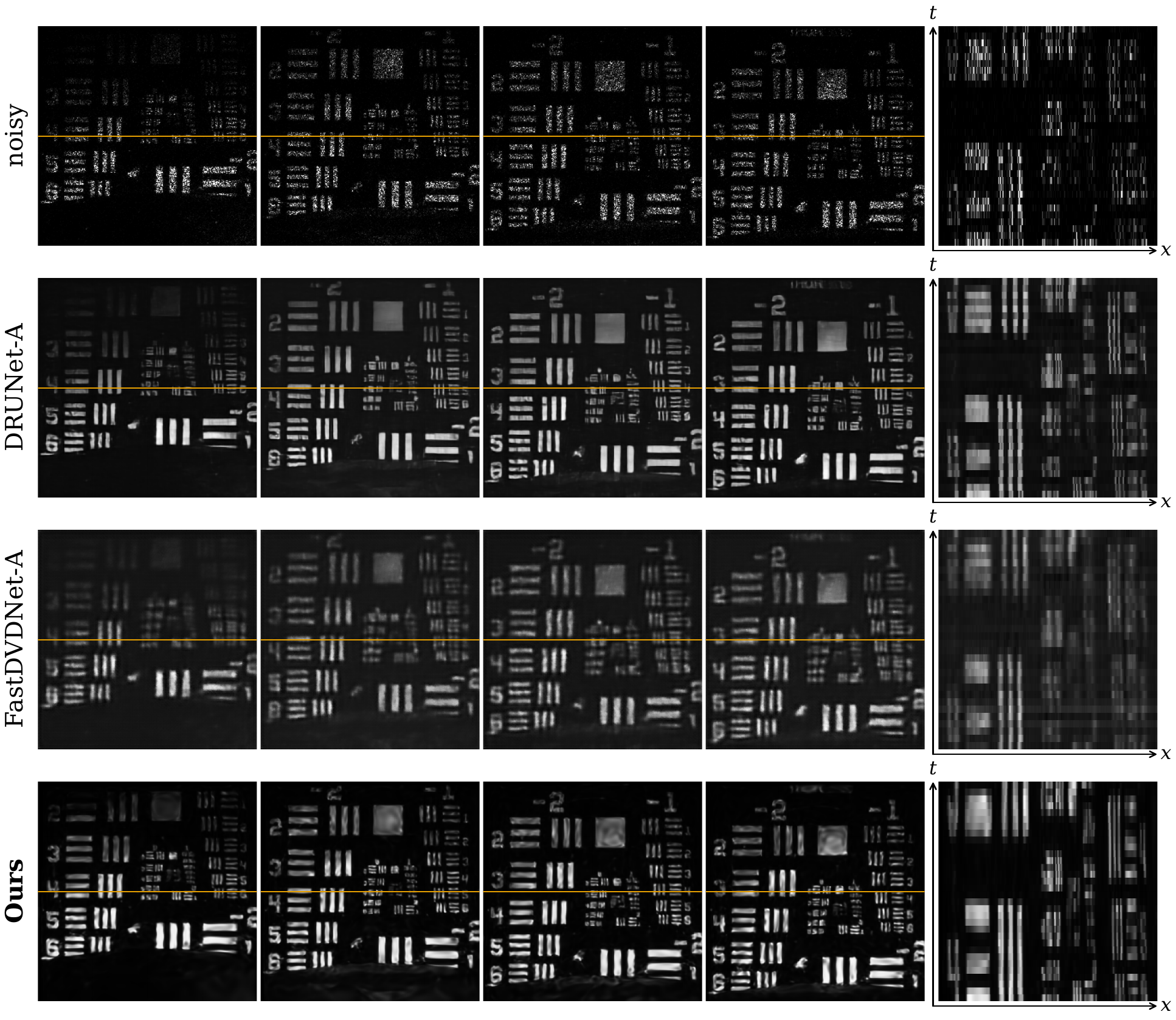}
    \caption{\textbf{Improved fine detail denoising on lab data.} Comparison of denoising methods on laboratory digital holography data of a translating resolution target. Our method resolves higher frequency spatial details while keeping them consistent frame-to-frame.}
    \label{fig:supp_lab_bar_chart}
\end{figure}

Figure~\ref{fig:supp_lab_bar_chart} shows full reconstruction results for the translating USAF resolution target summarized by the peak-profile plot in the main text. All methods substantially reduce visible speckle, and DRUNet-A in particular produces visually competitive frame-wise reconstructions. However, differences become apparent in the high-frequency resolution bars and in the $x$--$t$ slices. DRUNet-A recovers some fine structure, but several high-frequency resolution bars---particularly near the target center---merge together and vary from frame to frame. FastDVDNet-A produces smoother temporal behavior, but at the cost of stronger spatial blurring and reduced resolution of the finer bars. Our method preserves more high-frequency bar structure while maintaining consistent features across time.

\section{Sequence Length Dependence}
The temporal window length is a user-selected hyperparameter of the proposed method. Increasing the number of frames provides additional independent speckle realizations, which can improve denoising by giving the likelihood more information about the underlying reflectance. At the same time, longer temporal windows require the INR to represent a more complex spatiotemporal scene with fixed model capacity. Very short windows therefore provide limited speckle diversity, while very long windows can exceed the representational capacity of the INR and reduce reconstruction quality. The window length therefore controls a tradeoff between temporal diversity and the ability of the INR to compactly represent the dynamic scene.

\begin{table}[h]
\centering
\caption{\textbf{Effect of temporal window size.} Denoising performance for our method as a function of temporal window size. Results are shown for the DERF-HD \texttt{sunflower} sequence with simulated large-aperture speckle.}
\label{tab:num_frames_ablation}
\footnotesize
\setlength{\tabcolsep}{6pt}
\begin{tabular}{c|ccc}
\hline
Frames & PSNR $\uparrow$ & SSIM $\uparrow$ & ST-RRED $\downarrow$ \\
\hline
4  & 24.0 & 0.63 & 508 \\ 
8  & 25.8 & 0.74 & 401 \\ 
16 & 27.2 & 0.81 & 248 \\ 
32 & \textbf{27.9} & \textbf{0.84} & 147 \\ 
64 & 27.6 & \textbf{0.84} & \textbf{138} \\ 
\hline
\end{tabular}
\end{table}

To evaluate this tradeoff, we measured denoising performance as a function of temporal window length, with results shown in Table~\ref{tab:num_frames_ablation}. These comparisons use oracle early-stopping for all methods, rather than our early stopping rule, to allow for direct comparison. Performance improves substantially from 4 to 32 frames, indicating that the method benefits from additional temporal measurements. Increasing the window length to 64 frames slightly reduces PSNR, while maintaining SSIM and marginally improving ST-RRED. Based on this balance of spatial fidelity, temporal consistency, and computational cost, we use 32-frame fitting throughout the main paper. Larger temporal windows may lead to higher quality reconstructions with increased INR width or depth, but we chose to fix our network hyperparameters for this work.

\bibliography{supplement_refs}